\documentclass[a4paper,11pt]{article}
\pdfoutput=1 
\usepackage{jheppub} 
\usepackage[T1]{fontenc} 
\usepackage{amsmath}
\usepackage{slashed}
\usepackage{amsfonts}
\usepackage{amssymb}
\usepackage{mathtools}
\usepackage{multirow}
\usepackage{caption}
\usepackage{subcaption}
\usepackage{blindtext}
\usepackage{color}

\usepackage{graphicx}

\title{\textbf{$D^0\rightarrow \bar p e^+$ Form Factors in LCSR }}
\author[a,b,1]{Anshika Bansal,\note{Corresponding author.}}

\affiliation[a]{Physical Research Laboratory, Ahmedabad, 380009, India.}
\affiliation[b]{Indian Institute of Technology, Gandhinagar, 382424, India.}

\emailAdd{anshika@prl.res.in}

\abstract{Baryon number is conserved in the Standard Model (SM) of particle physics. Baryon Number Violation (BNV) is one of the criteria to explain the matter anti-matter asymmetry of the universe. Various beyond the SM (BSM) scenarios motivate BNV. In that view, it becomes important to look for various BNV decays. \\
In this work, we consider the BNV decay of heavy charmed meson to anti-proton and positron. This decay proceeds via the dim-6 SM effective field theory (SMEFT) operator and involves 12 independent form factors. We give estimates for these form factors in the framework of Light Cone Sum Rules (LCSR). As the interpolation current for the proton state is not unique, we have considered different forms of proton interpolation current and discuss the effect of the choice of the interpolation current.  }

\begin{document} 
\maketitle
\flushbottom
\section{Introduction}
\label{intro}
According to Sakharov's conditions \cite{Sakharov:1967dj}, baryon number violation (BNV) is an important criterion to explain the matter-anti-matter asymmetry of the universe. But the baryon number turns out to be an accidental symmetry in the Standard Model (SM) of particle physics. This leads us to look for the physics beyond the SM (BSM). In order to do that, one can look for the BNV processes at the experimental facilities as they will be a clear signature of new physics. Till date, we have not observed any BNV process, but the experiments put very stringent bounds on the decay widths of several BNV processes like proton decay, decays of heavy mesons to baryons, etc \cite{ParticleDataGroup:2020ssz}. Theoretically, BNV processes are well motivated in various BSM theories like Grand Unification Theories (GUTs), super Symmetry (SUSY), models of baryogenesis, etc (see for eg. \cite{Pati:1973uk,Georgi:1974sy,Dienes:1998vh,Nilles:1983ge,Morrissey:2012db,FileviezPerez:2015mlm}, and reference therein). In these scenarios, the BNV processes are possible via the exchange of heavy gauge vector or scalar bosons like leptoquarks. In the framework of SM effective field theory (SMEFT), these processes can be studies using BNV higher dimension operators which can be obtained by integrating out these heavy particles. Out of all the BNV processes, proton decay has got the most attention so far, theoretically (see for eg. \cite{Aoki:2017puj,Haisch:2021hvj,Bansal:2022xbg} and references therein) as well as experimentally (see for eg. \cite{Super-Kamiokande:2020tor,Dev:2022jbf} and references therein) . With the advances in the experimental facilities, it becomes important that one pays proper theoretical attention to other processes as well. Very recently, BESIII collaboration updated the upper limit on the branching fraction of $D^0\rightarrow \bar p e^+$ to be $<1.2 \times 10^{-6}$ \cite{BESIII:2021krj}. To the best of our knowledge, there are no theoretical estimates for this decay mode so far. In order to get the theoretical estimates, one needs an input for the form factors (FFs) involved in this process.
In the present work, we attempt to get an estimate for these FFs using the method of Light Cone Sum Rules (LCSR). \\
The rest of the paper is organised as follows; In Section-2, we describe the general parameterisation of the decay amplitude for the process in terms of the FFs. In Section-\ref{FF}, we derive these FFs in the framework of LCSR and provide the numerical analysis in Section-\ref{NA}. Finally, we conclude our findings and discuss the results in Section-\ref{CD}. This article includes 4 Appendices. In Appendix-\ref{A}, we provide the useful identities and integrals involved. In Appendix-\ref{C}, we provide the analytic expressions of the correlation functions involved for a general interpolation current for the proton state. Finally, we collect the numerical values of all the parameters involved in the calculations in Appendix-\ref{D}. 
\section{Amplitude parametrisation}
\label{Ampp}
As discussed above, BNV processes can be calculated in the BSM scenarios using the higher dimensional effective operators. In the SMEFT, there are 4 types of dimension-6 operators which can lead to BNV processes. These operators respect the SM gauge group but violate the baryon number which is an accidental symmetry of the SM. These BNV operators can be written as \cite{Weinberg:1979sa, Wilczek:1979hc, Abbott:1980zj}
\begin{align}
 &\mathcal{O}_{ijkl}^{duq\ell}=\epsilon_{abc}\epsilon_{\alpha\beta}\left(d_i^a C u_j^b\right)\left(q_k^{\alpha c } C \ell_l^\beta\right), \hspace{1.8cm}\mathcal{O}_{ijkl}^{qque}= \epsilon_{abc}\epsilon_{\alpha\beta}\left(q_i^{\alpha a} C q_j^{\beta b}\right)\left(u_k^{ c } C e_l\right) \nonumber \\ 
 &\mathcal{O}_{ijkl}^{qqq\ell}=\epsilon_{abc}\epsilon_{\alpha\beta}\epsilon_{\gamma\delta}\left(q_i^{\alpha a} C q_j^{beta b}\right)\left(q_k^{\gamma c } C \ell_l^\delta\right), \hspace{0.8cm}\mathcal{O}_{ijkl}^{duue}= \epsilon_{abc}\left(d_i^{a} C u_j^{ b}\right)\left(u_k^{ c } C e_l\right)
\end{align}
Here, $\{i,j,k,l\}$ represent the flavour indices, $\{a,b,c\}$ are the color indices, $\{u,d\}$ represent the right-handed up and down quarks and $\{q,\ell\}$ represent the left-hand doublets of quarks and leptons. The Einstein's convention of summation over repeated indices is adopted.
Using these operators and the generalised Fierz transformations \cite{Nieves:2003in}, the baryon number violating lagrangian which will contribute to $D^0\rightarrow \bar p e^+$ can be written as,
\begin{equation}
  \mathcal{L}_{\slashed B}^{(6)}=\sum_{\Gamma,\Gamma'} c_{\Gamma\Gamma'}^A\mathcal{O}_{\Gamma\Gamma'}^A=\sum_{\Gamma\Gamma'} c_{\Gamma\Gamma'}^A \epsilon^{ijk}\left(d^T_i C P_{\Gamma}  \Gamma_A u_j\right)\left(e^T C P_{\Gamma'}\Gamma^A  c_k\right)
  \label{lag}
\end{equation}
Here, $C=i \gamma^2\gamma^0$ is the charge conjugation matrix, superscript $T$ represents the transpose, $P_{\Gamma}$ and $P_{\Gamma'}$ are the chirality projection operators with $\{\Gamma,\Gamma'\}\in \{L,R\}$ and $\Gamma^A \in \{1,\gamma_\mu,\sigma^{\mu\nu}\}$ with $A\in\{S,V,T\}$. $c_{\Gamma\Gamma'}^A$ are the Wilson coefficients \footnote{The Lagrangian is assumed to be extracted in terms of the physical fields at the charm scale and thus, $c_{\Gamma\Gamma'}^A$'s also include all the flavour and usual RG running effects.}. The transition amplitude for $D^0\rightarrow \bar p e^+$ is defined as the matrix element of this Lagrangian between the initial and the final states as
 \begin{align}
  \mathcal{A}(D^0(p_D)\rightarrow \bar p(p_p) e^+(p_e))&= \sum_{\Gamma,\Gamma'} c_{\Gamma\Gamma'}^A \left<e^+(p_e)\bar p(p_p)\left|\mathcal{O}_{\Gamma\Gamma'}^A\right|D^0(p_D)\right> 
  \label{amp}
  \end{align}
The leptonic and the hadronic parts can be factorised here such that,
\begin{equation}
    \mathcal{A}(D^0(p_D)\rightarrow \bar p(p_p) e^+(p_e))= \sum_{\Gamma,\Gamma'} c_{\Gamma\Gamma'}^A \bar v_e^c H_{\Gamma\Gamma'}^A v_p(p_p)
    \end{equation}
where, $\bar v_e^c$ is the spinor corresponding to the positron and $H_{\Gamma\Gamma'}^A v_p(p_p)$ is the hadronic object of interest and is given by
\begin{equation}
\label{had}
    H_{\Gamma\Gamma'}^A v_p(p_p) = \left<\bar p(p_p)\left|\epsilon^{ijk}\left(d^T_i C \Gamma_A P_{\Gamma} u_j\right)\left(\Gamma^A P_{\Gamma'} c_k\right)\right|D^0(p_D)\right>
\end{equation}
This hadronic object can be generally parameterized as,
\begin{equation}
 H_{\Gamma\Gamma'}^A v_p(p_p)= P_{\Gamma'} \left(F_{\Gamma\Gamma'}^{A,0}(p_e^2)+ \slashed v F_{\Gamma\Gamma'}^{A,1}(p_e^2)\right)v_p(p_p)
\end{equation}
Here, $F_{\Gamma\Gamma'}^{A,n}(p_e^2)$ are the form factors with $A\in\{S,V,T\}$ and $n\in\{0,1\}$. As D-meson comprises of a heavy quark, we will treat it in the framework of heavy quark effective theory (HQET) \cite{Politzer:1988bs,Manohar:2000dt}. $v$ is the velocity of the D-meson such that $p_D^\mu = m_D v^\mu $ with $v^2=1$.\\
The parity conservation in QCD leads to the following relations between these FFs,
\begin{align}
 & F_{LL}^{A,n}= F_{RR}^{A,n} \hspace{2cm }  F_{LR}^{A,n}= F_{RL}^{A,n}
\end{align}
Hence, there are effectively 12 independent FFs. 
We will compute these FFs in the framework of the light cone sum rules (LCSR).

\section{Form Factors in LCSR}
\label{FF}
LCSR is a technique to compute the hadronic objects of interest using the analytic properties of the correlation function involved in the process. Any LCSR computation involves four basic tools: dispersion relation, light cone operator product expansion, quark hadron duality and Borel transformation. We will not discuss them in detail here and suggest the interested readers to refer to \cite{Colangelo:2000dp,Khodjamirian:2020btr, Shifman:1978bx, Shifman:1998rb}.\\
To obtain the correlation function involved here, we first need to opt for an interpolation current for the anti-proton state. The interpolation current for the proton is not unique \cite{Ioffe:1982ce,Leinweber:1994nm}. We choose a general form for the interpolation current as 
\begin{equation}
    \chi(x) = \chi_1(x)+t \chi_2(x)
    \label{int}
\end{equation}
where, $\chi_1(x)$ and $\chi_2(x)$ are defined as,
\begin{equation}
 \chi_1(x)= \epsilon^{lmn}\left(u^T_l(x) C \gamma_5 d_{m}(x)\right)u_n(x), \hspace {1cm}\chi_2(x)= \epsilon^{lmn}\left(u^T_l(x) C  d_{m}(x)\right)\gamma_5u_n(x)
\end{equation}
where, $\{l,m,n\}$ are the color indices. The interpolation current is defined such that $\langle\bar p(p_p) | \chi(0)|0\rangle = m_p\lambda_p v_p(p_p)$ where, $\lambda_p$ is a measure of the strength with which this current couples with the proton/antiproton state.\\
We interpolate the proton state using this current in Eq.(\ref{had}) and get the relevant correlation function which read as
\begin{align}
 \Pi_{\Gamma\Gamma'}^A&= i \int d^4 x \hspace{0.1cm} e^{ip_e.x}\left<0\left|T\{\bar \chi(0)  \mathcal{Q}_{\Gamma\Gamma'}^A(x)\}\right|D^0(v)\right> 
\end{align}
where,  $T$ denotes the time ordering,  $\mathcal{Q}_{\Gamma\Gamma'}^A(x)=\epsilon^{ijk}\left(d^T_i C  P_{\Gamma}\Gamma_A u_j\right)\left(P_{\Gamma'}\Gamma^A  c_k\right)$ and $\bar \chi(0) = \chi^\dagger \gamma_0$. According to the matching condition of the LCSR, 
\begin{equation}
    \Pi_{\Gamma\Gamma'}^{A,had} (p_p,p_e) = \Pi_{\Gamma\Gamma'}^{A,QCD}(p_p,p_e)
\end{equation}
where, $\Pi_{\Gamma\Gamma'}^{A,had} (p_p,p_e)$ and $\Pi_{\Gamma\Gamma'}^{A,QCD}(p_p,p_e)$ are the hadronic and quantum chromodynamics (QCD) parameterizations of the correlation function, respectively. Here, the hadronic parameterization can be obtained by inserting the complete set of intermediate states. Once we separate the pole contribution coming from the proton state, we get,
\begin{align}
    \Pi_{\Gamma\Gamma'}^{A,had} &= - m_p \lambda_p \bar v_p(p_p)\left[H_{\Gamma\Gamma'}\right]v_p(p_p)+\ldots \nonumber \\
    &= iP_{\Gamma'}\left[\Pi_{\Gamma\Gamma'}^{A,S}(p_p^2,p_e^2)+\Pi_{\Gamma\Gamma'}^{A,V}(p_p^2,p_e^2) \slashed v+\Pi_{\Gamma\Gamma'}^{A,P}(p_p^2,p_e^2) \frac{\slashed p_p}{m_p}+\Pi_{\Gamma\Gamma'}^{A,VP}(p_p^2,p_e^2) \frac{\slashed v \slashed p_p}{m_p}\right]
    \label{corfn}
\end{align}
The contributions coming from the continuum and the heavier states are represented here by ellipses. $\Pi_{\Gamma\Gamma'}^{A,r}$ are the scalar function of $p_p^2$ and $P_e^2=-p_e^2$ with $r=\{S,V,P,VP\}$. Using the dispersion relation, these functions can be written in terms of the spectral densities which can inturn be related to the imaginary part of the correlation function (for details look at \cite{Bansal:2022xbg}). Separating the ground state pole contribution coming from the proton state and the continuum and heavier states contributions, these spectral densities can be written as
\begin{equation}
     \rho_{\Gamma\Gamma'}^{A,r,had}(s,P_e^2)= \lambda_p m_p^2 \delta(s-m_p^2) F_{\Gamma\Gamma'}^{A,r}(s,P_e^2)+ \frac{1}{\pi} \text{Im}\left(\Pi_{\Gamma\Gamma'}^{A,r,had}(s,P_e^2)\right)
\end{equation}
Here, $F_{\Gamma\Gamma'}^{A,r}(s,P_e^2)$ are related to $F_{\Gamma\Gamma'}^{A,n}(s,P_e^2)$ for the on-shell proton, i.e. $s=m_p^2$, as
\begin{align}
    &-F_{\Gamma\Gamma'}^{A,S} = F_{\Gamma\Gamma'}^{A,P} = F_{\Gamma\Gamma'}^{A,0} \nonumber \\
     &-F_{\Gamma\Gamma'}^{A,V} = F_{\Gamma\Gamma'}^{A,VP} = F_{\Gamma\Gamma'}^{A,1}
\end{align}
The final dispersion relation for the correlation function in Eq.(\ref{corfn}) reads as
\begin{equation}
    \Pi_{\Gamma\Gamma'}^{A,r,had}(p_p^2,P_e^2) = \lambda_p m_p^2 \frac{F_{\Gamma\Gamma'}^{A,r}}{m_p^2-p_p^2} + \int_{s_0^h}^\infty ds \frac{1}{\pi} \frac{\text{Im}\left(\Pi_{\Gamma\Gamma'}^{A,r,had}(s,P_e^2)\right)}{s-p_p^2}
\end{equation}
where, $s_0^h$ is the continuum threshold.\\
In order to get the QCD parameterization of the correlation function, we first need the time ordered product of $\mathcal{Q}_{\Gamma\Gamma'}^A(x)$ and $\bar \chi(0)$ which is given by
\begin{align}
    T\left\{\bar \chi(0)  \mathcal{Q}_{\Gamma\Gamma'}^A(x)\right\}= -\epsilon^{lmn}\epsilon^{ijk}&\left[\left(P_{\Gamma'}\Gamma^A  c_i(x)\right)\left\{ \left(\bar u_l(0) \gamma_5 \tilde  S_{mj}^d(x) P_\Gamma \Gamma_A S_{nk}^u(x) \nonumber \right. \right.  \right.\\ & \left.\left. \left.+ \bar u_l(0) \text{Tr}\left(\gamma_5 \tilde S_{mj}^d(x) P_{\Gamma}\Gamma_A  S_{nk}^u(x)\right)\right) \nonumber \right. \right. \\ & \left. \left. +t\left( \bar u_l(0) \tilde  S_{mj}^d(x) P_\Gamma \Gamma_A S_{nk}^u(x)\gamma_5\nonumber \right. \right.\right. \\ & \left.\left.\left.+ \bar u_l(0)\gamma_5  \text{Tr}\left(\tilde S_{mj}^d(x) P_{\Gamma}\Gamma_A  S_{nk}^u(x)\right) \right)\right\}\right]
\end{align}
Here,  $\tilde \Gamma = C\Gamma C^{-1}$ and $S_{ij}^q(x)$ is the quark propagator at the light like separations and is given by
\begin{equation}
    S_{ij}^q(x)= \left(\frac{i \slashed x}{2 \pi^2 x^4} -\frac{ \left<\bar q q\right>}{12}\right)\delta_{ij}+\ldots
\end{equation}
considering quark to be mass-less. $\left<\bar q q\right>$ represents the quark condensates and the ellipses represents the higher terms involving one or more gluons and are not considered in the present analysis. \\
Secondly, we need the matrix element of the quark bilinear which can be defined in terms of the distribution amplitudes of D-meson as \cite{Beneke:2004dp}
\begin{equation}
    \left<0\left|\bar u_\alpha(0) [x,0] c_\beta(x)\right|D^0(v)\right> = \frac{-i f_D m_D}{4} \int_0^\infty d w \hspace{0.1cm} e^{iwv.x} \left[(1+\slashed v)\left\{\phi_+^D(w)-\frac{\phi_+^D(w)-\phi_-^D(w)}{2v.x}\slashed x\right\}\gamma_5\right]_{\beta\alpha}
\end{equation}
Here, $f_D$ is the decay constant of D-meson, $\phi_+^D(w)$ and $\phi_-^D(w)$ are the light cone distribution amplitudes (LCDAs) of D-meson. We use the exponential mode parameterization for the LCDAs of the D-meson \cite{Grozin:1996pq} which reads as
\begin{equation}
    \phi_D^+(w) = \frac{1}{w_0^2} e^{-w/w_0}, \hspace{2 cm}  \phi_D^-(w) = \frac{1}{w_0} e^{-w/w_0}
\end{equation}
where, $w_0$ is a model input parameter.\\
Using the above definitions and the integrals collected in Appendix-\ref{A}, the correlation function in QCD reads as,
\begin{align}
    \Pi_{\Gamma\Gamma'}^{A,QCD}
    &= iP_{\Gamma'}\left[\Pi_{\Gamma\Gamma'}^{A,S,QCD}(p_p^2,p_e^2)+\Pi_{\Gamma\Gamma'}^{A,V}(p_p^2,p_e^2) \slashed v+\Pi_{\Gamma\Gamma'}^{A,P}(p_p^2,p_e^2) \frac{\slashed p_p}{m_p}+\Pi_{\Gamma\Gamma'}^{A,VP}(p_p^2,p_e^2) \frac{\slashed v \slashed p_p}{m_p}\right]
\end{align}
Here, $\Pi_{\Gamma\Gamma'}^{A,S,QCD}(p_p^2,p_e^2)$ with $r=\{S,V,P,VP\}$ are the analytic functions of $p_p^2$ and $p_e^2$. The expressions for these functions are rather lengthy and hence we collect then in Appendix-\ref{C} for both $\Gamma\Gamma' = LL$ and $\Gamma\Gamma'=LR$ cases. \\
Now to derive the sum rules, we make use of the quark hadron duality to write the contribution of continuum and heavier states in the spectral density in terms of QCD calculated spectral functions as,
\begin{equation}
 \int_{s_0^h}^\infty ds \frac{1}{\pi} \frac{\text{Im}\left(\Pi_{\Gamma\Gamma'}^{A,r,had}(s,P_e^2)\right)}{s-p_p^2} \approx     \int_{s_0}^\infty ds \frac{1}{\pi} \frac{\text{Im}\left(\Pi_{\Gamma\Gamma'}^{A,r,QCD}(s,P_e^2)\right)}{s-p_p^2}
\end{equation}
where, $s_0$ is the continuum threshold and is not necessarily equals to $s_0^h$. It is a free parameter in the sum rule calculation. Finally, to get the final sum rule, we put all the things together and perform the Borel transformation to suppress the effect of continuum and heavier states. The statement of final sum rule reads as,
\begin{equation}
    F_{\Gamma\Gamma'}(P_e^2,s_0,M^2)= -\frac{e^{\frac{m_p^2}{M^2}}}{m_p^2 \lambda_p}\int_0^{s_0}ds e^{\frac{-s}{M^2}} \frac{1}{\pi} \text{Im}\left(\Pi_{\Gamma\Gamma'}^{A,r,QCD}(s,P_e^2)\right)
\end{equation}
where, $M$ is the Borel mass and is another free parameter in the sum rule calculations. We will discuss how to choose the values of $M$ and $s_0$ in the next section. 
\section{Results}
\label{NA}
\begin{figure}[h]
\centering
   \begin{subfigure}{0.45\textwidth}
    \centering
    \includegraphics[width=0.8\linewidth]{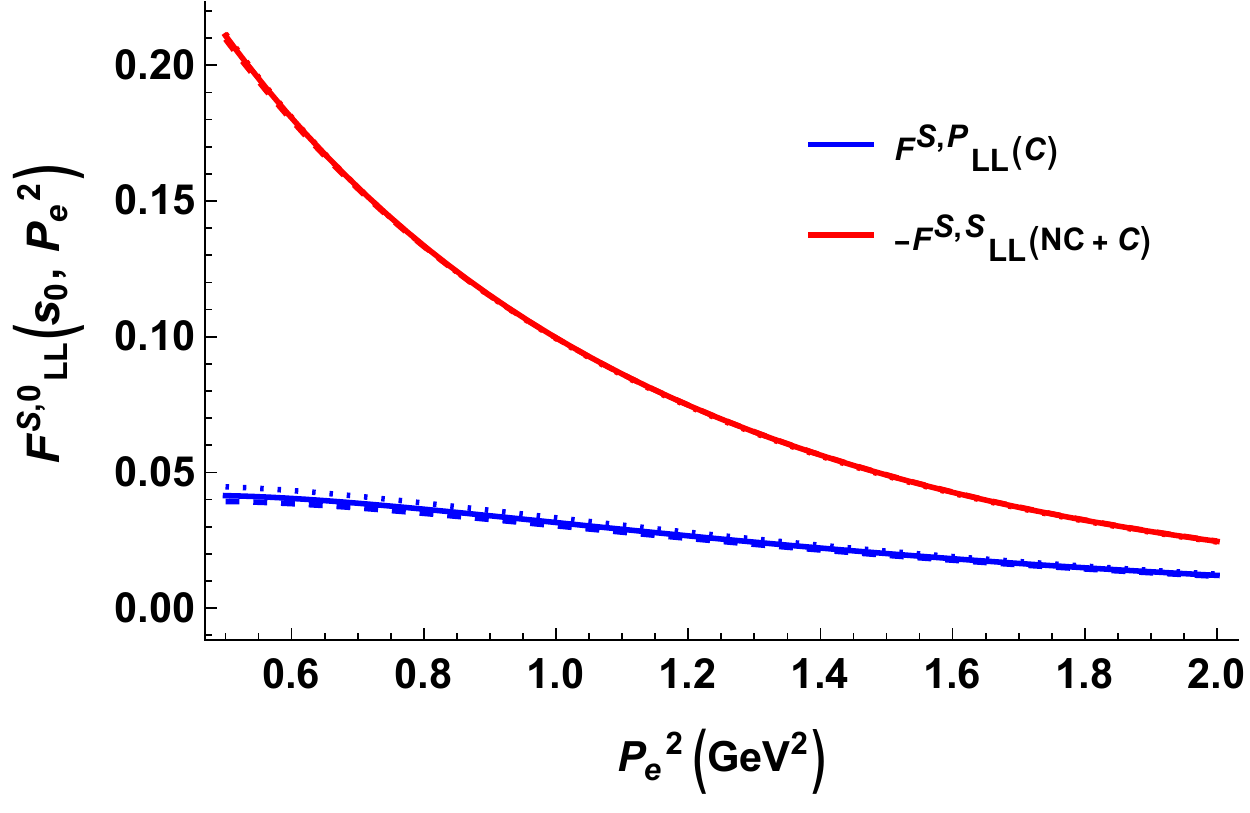}
   \end{subfigure}
   \begin{subfigure}{0.45\textwidth}
    \centering
    \includegraphics[width=0.8\linewidth]{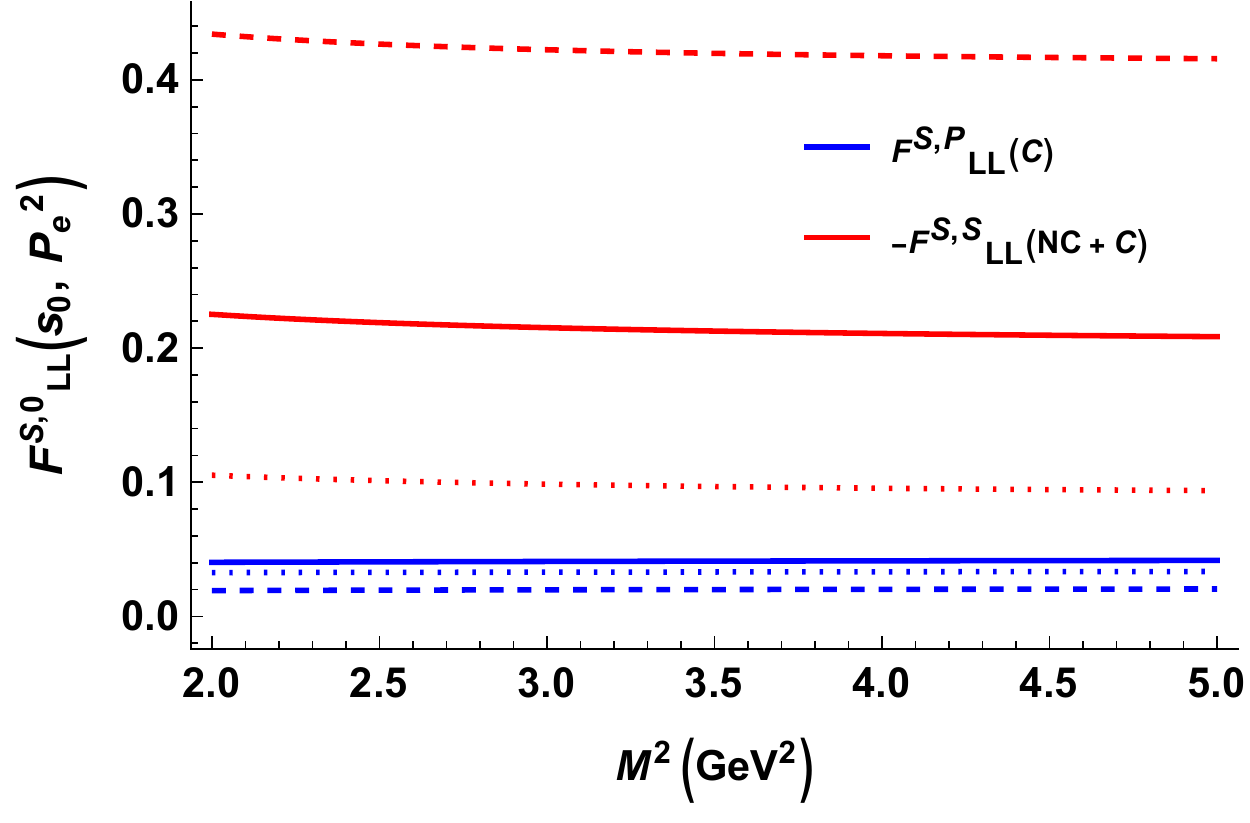}
   \end{subfigure}
   \begin{subfigure}{0.45\textwidth}
    \centering
    \includegraphics[width=0.8\linewidth]{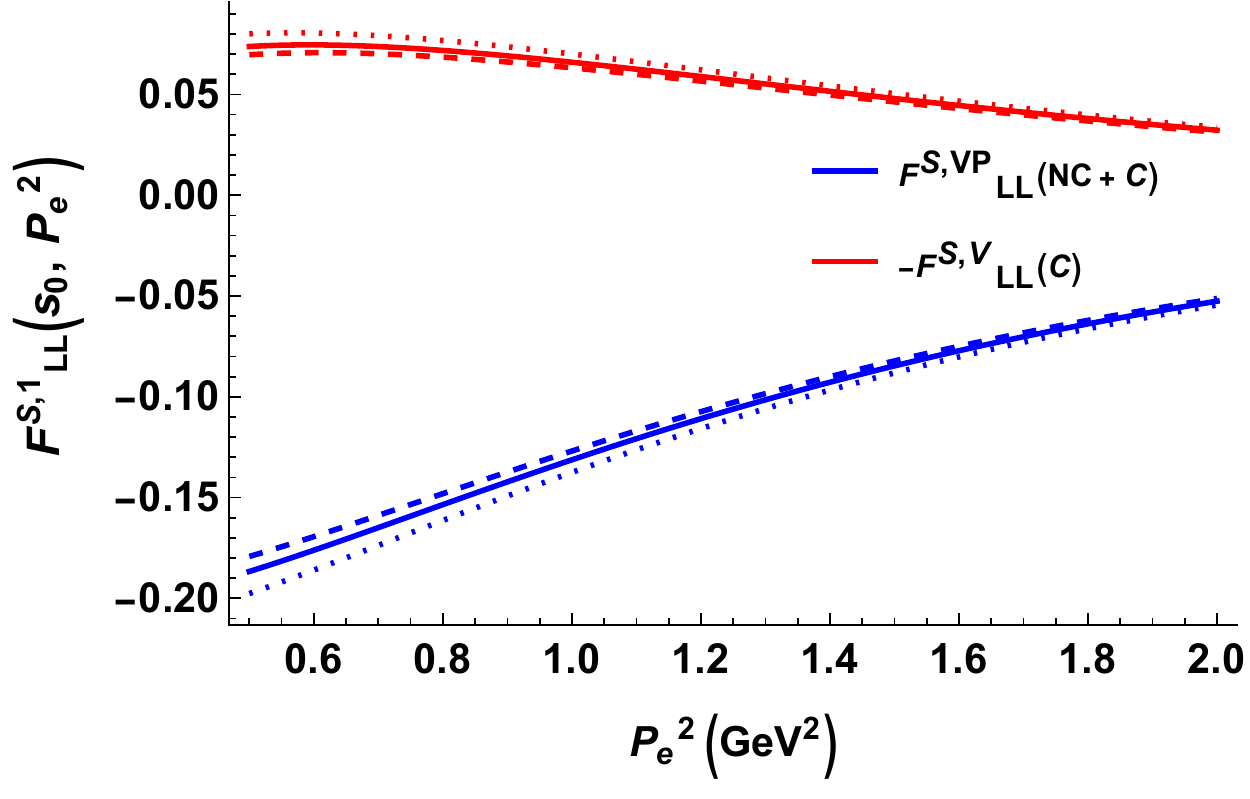}
   \end{subfigure}
   \begin{subfigure}{0.45\textwidth}
    \centering
    \includegraphics[width=0.8\linewidth]{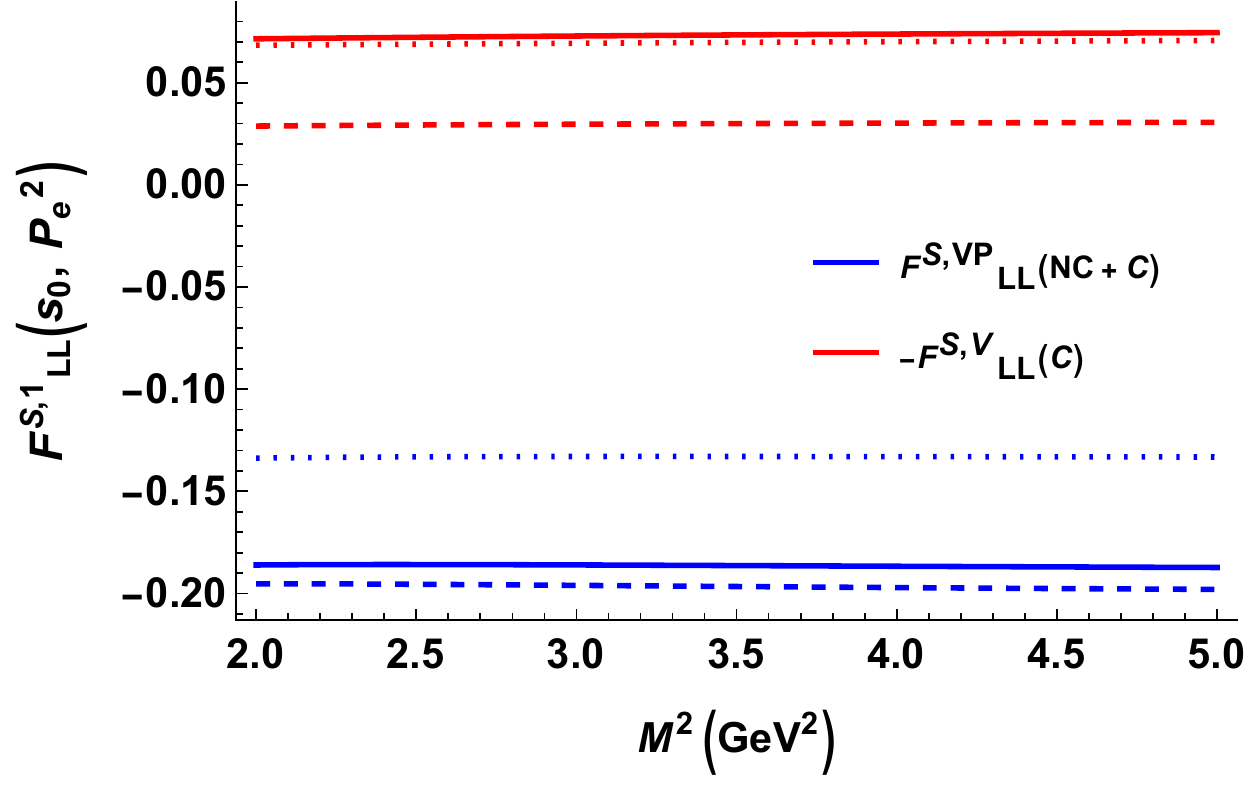}
   \end{subfigure}
   \begin{subfigure}{0.45\textwidth}
    \centering
    \includegraphics[width=0.8\linewidth]{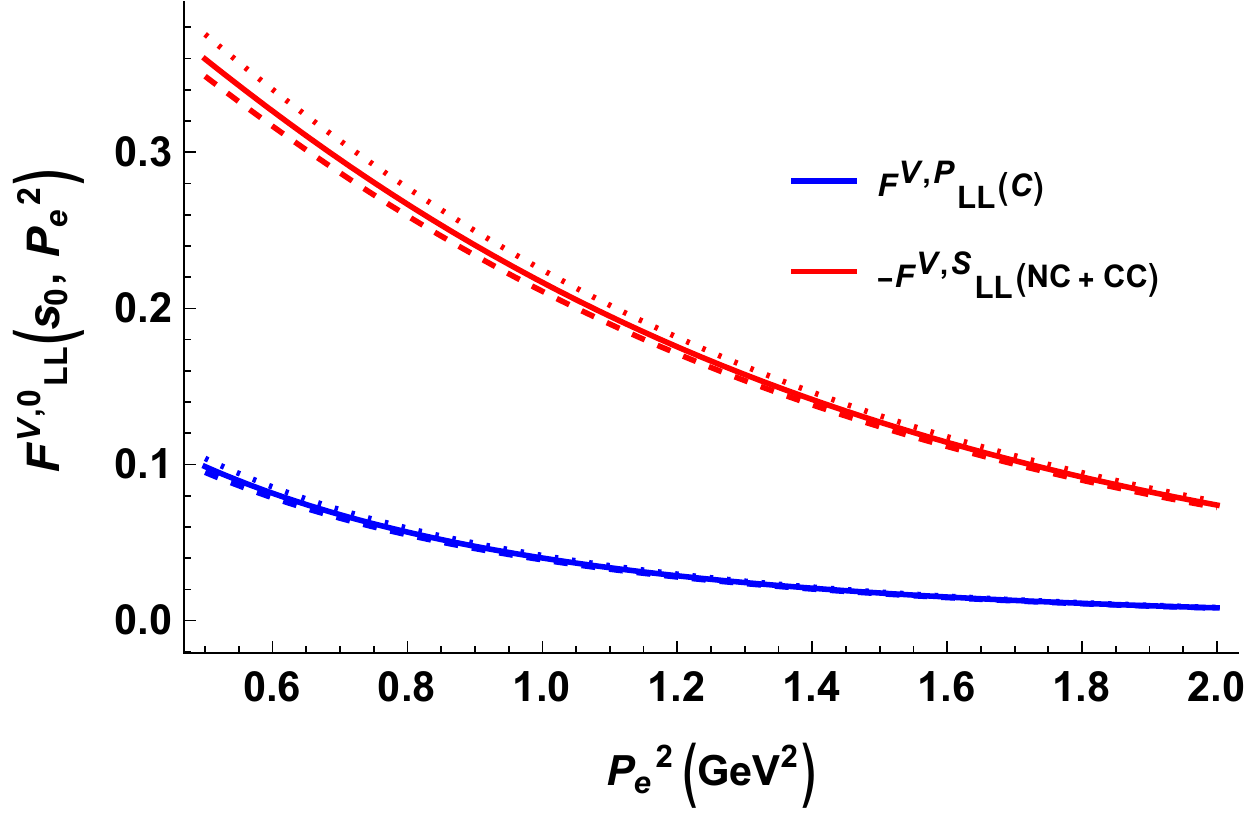}
   \end{subfigure}
   \begin{subfigure}{0.45\textwidth}
    \centering
    \includegraphics[width=0.8\linewidth]{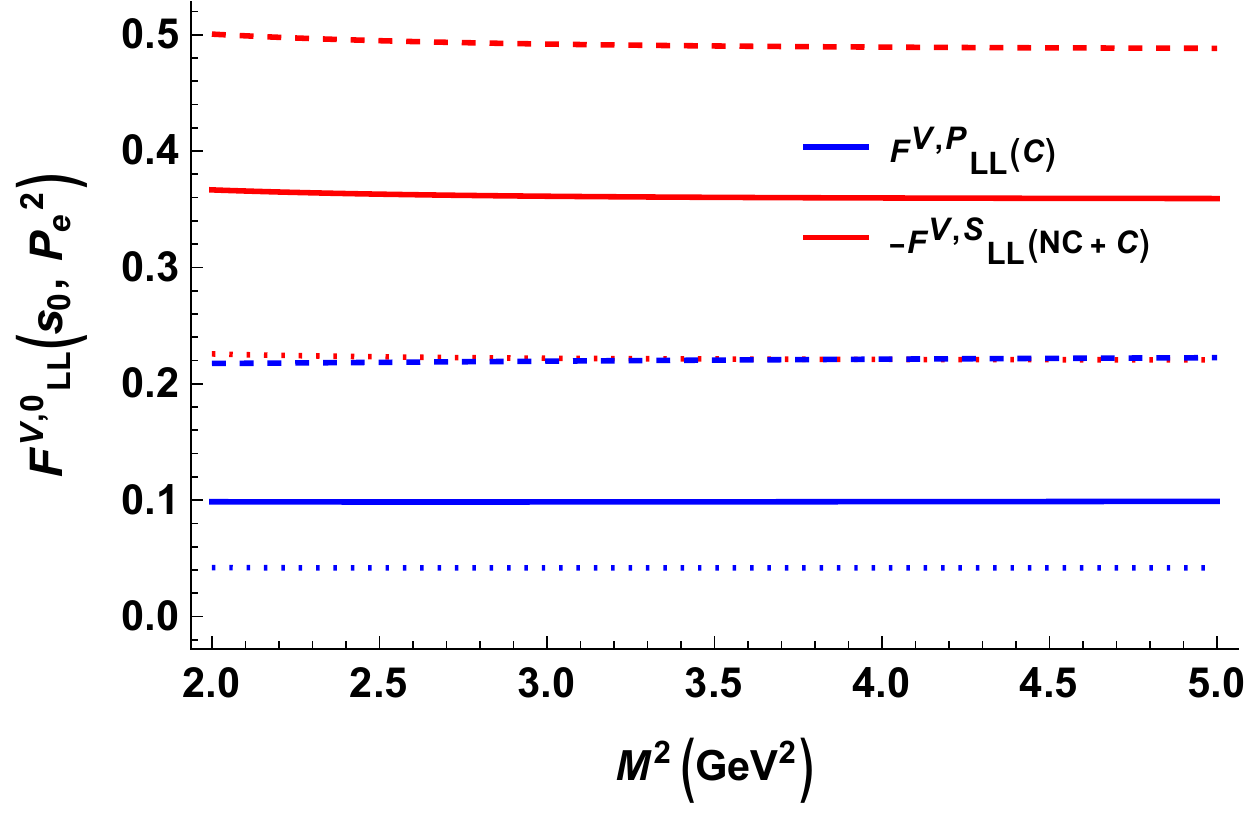}
   \end{subfigure}
   \begin{subfigure}{0.45\textwidth}
    \centering
    \includegraphics[width=0.8\linewidth]{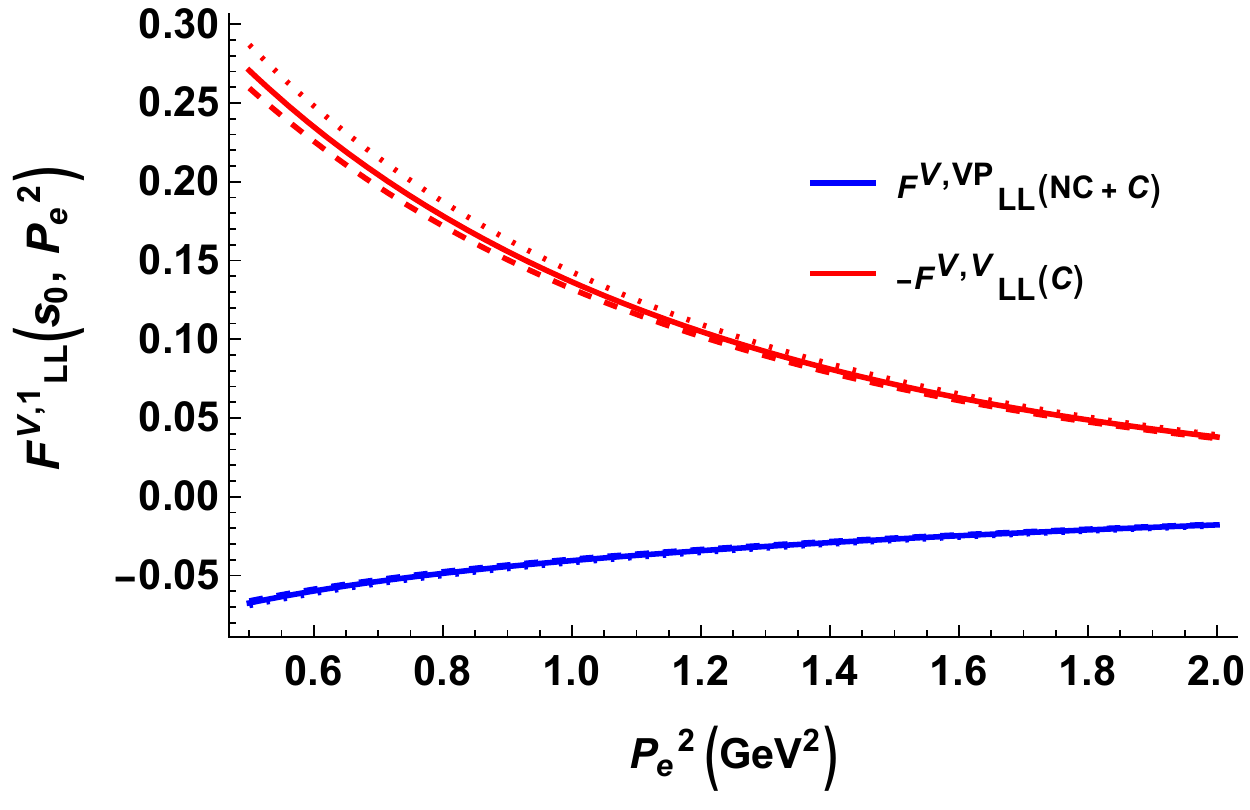}
   \end{subfigure}
   \begin{subfigure}{0.45\textwidth}
    \centering
    \includegraphics[width=0.8\linewidth]{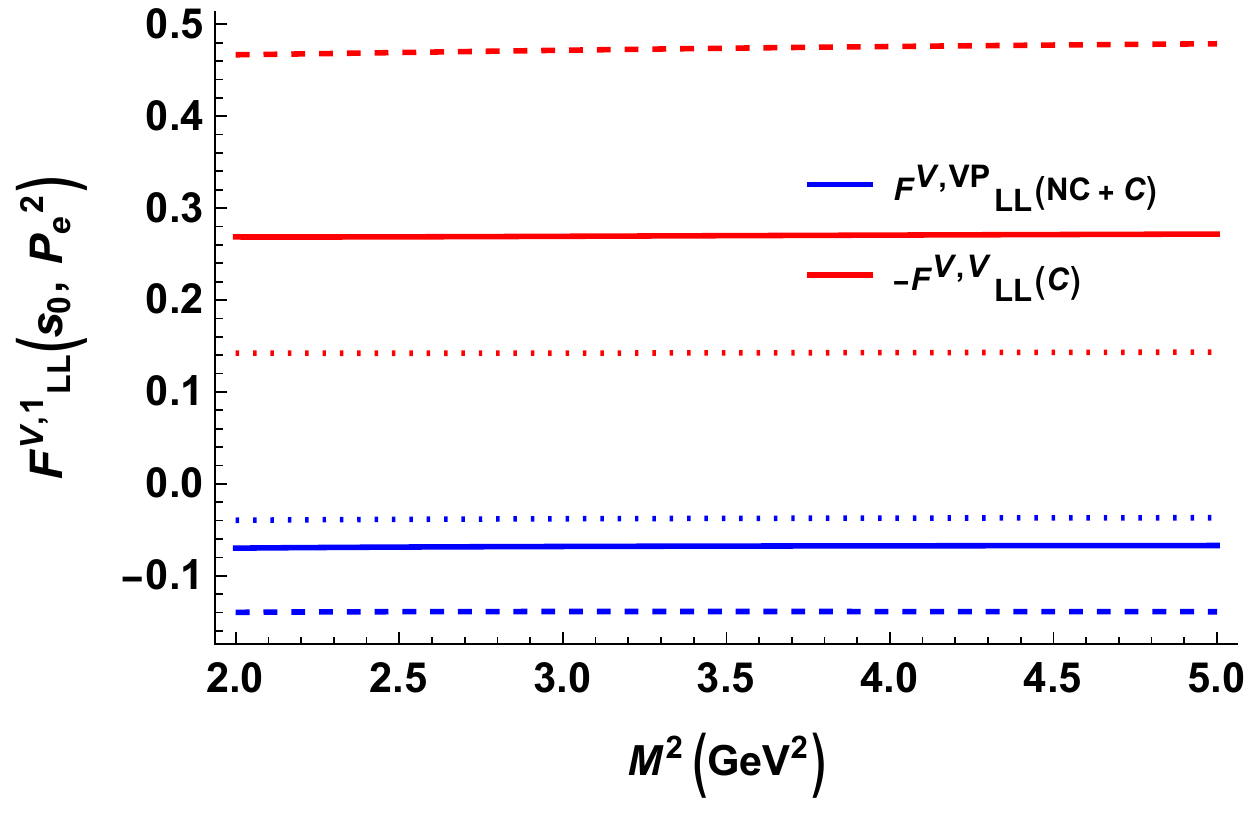}
   \end{subfigure}
   \begin{subfigure}{0.45\textwidth}
    \centering
    \includegraphics[width=0.8\linewidth]{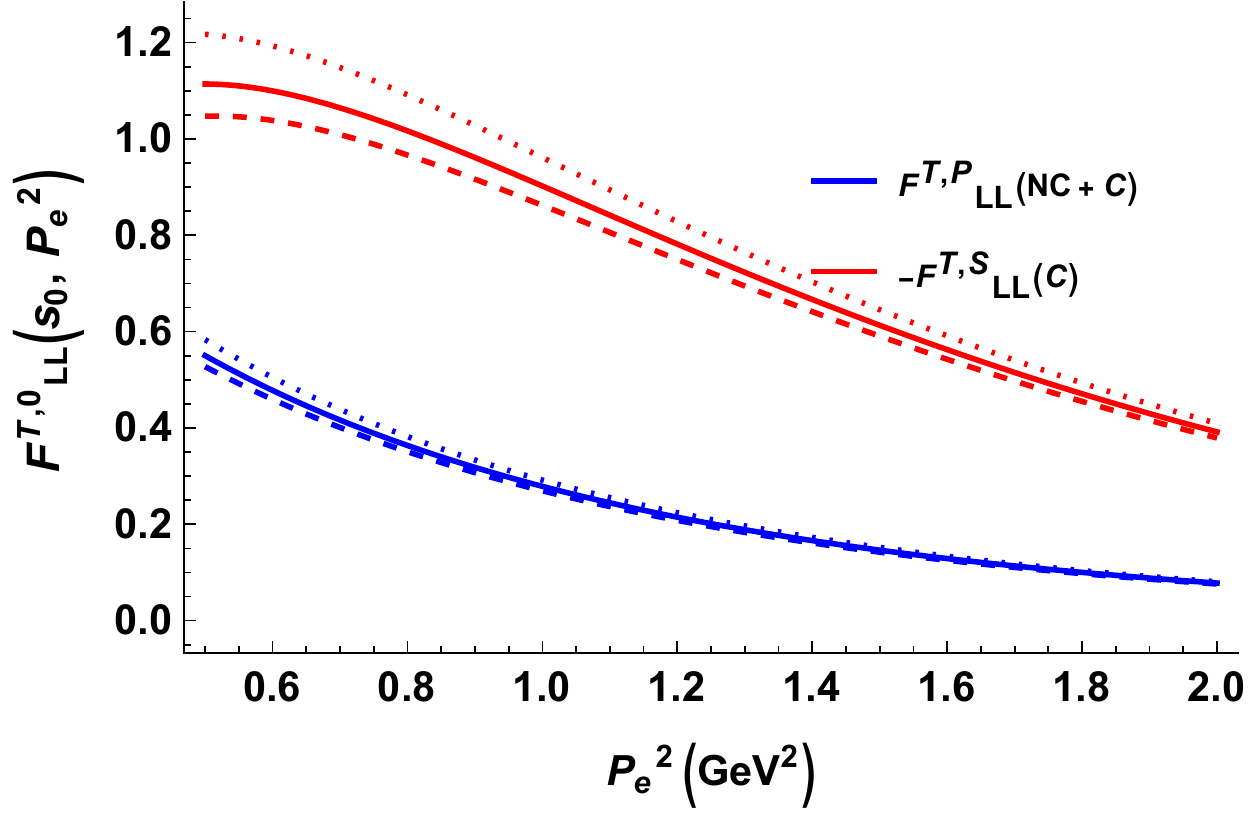}
   \end{subfigure}
   \begin{subfigure}{0.45\textwidth}
    \centering
    \includegraphics[width=0.8\linewidth]{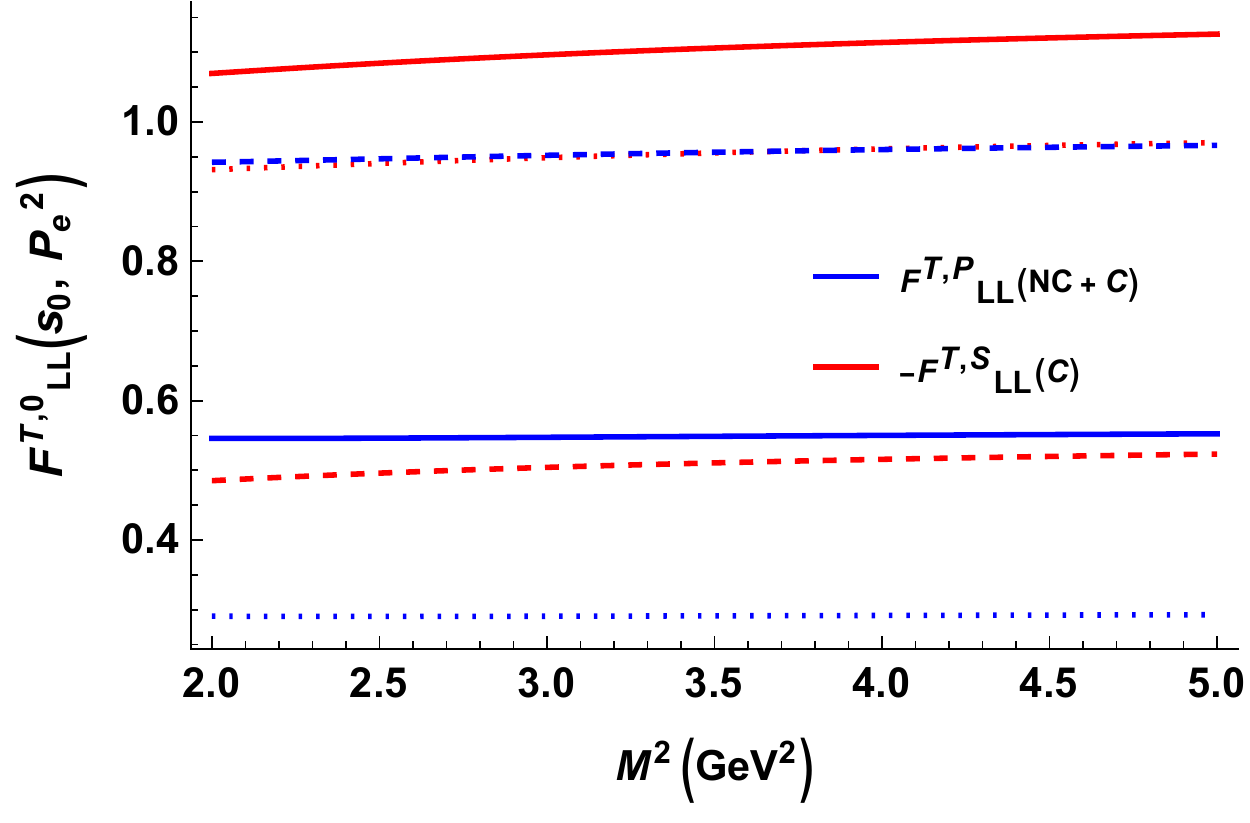}
   \end{subfigure}
   \begin{subfigure}{0.45\textwidth}
    \centering
    \includegraphics[width=0.8\linewidth]{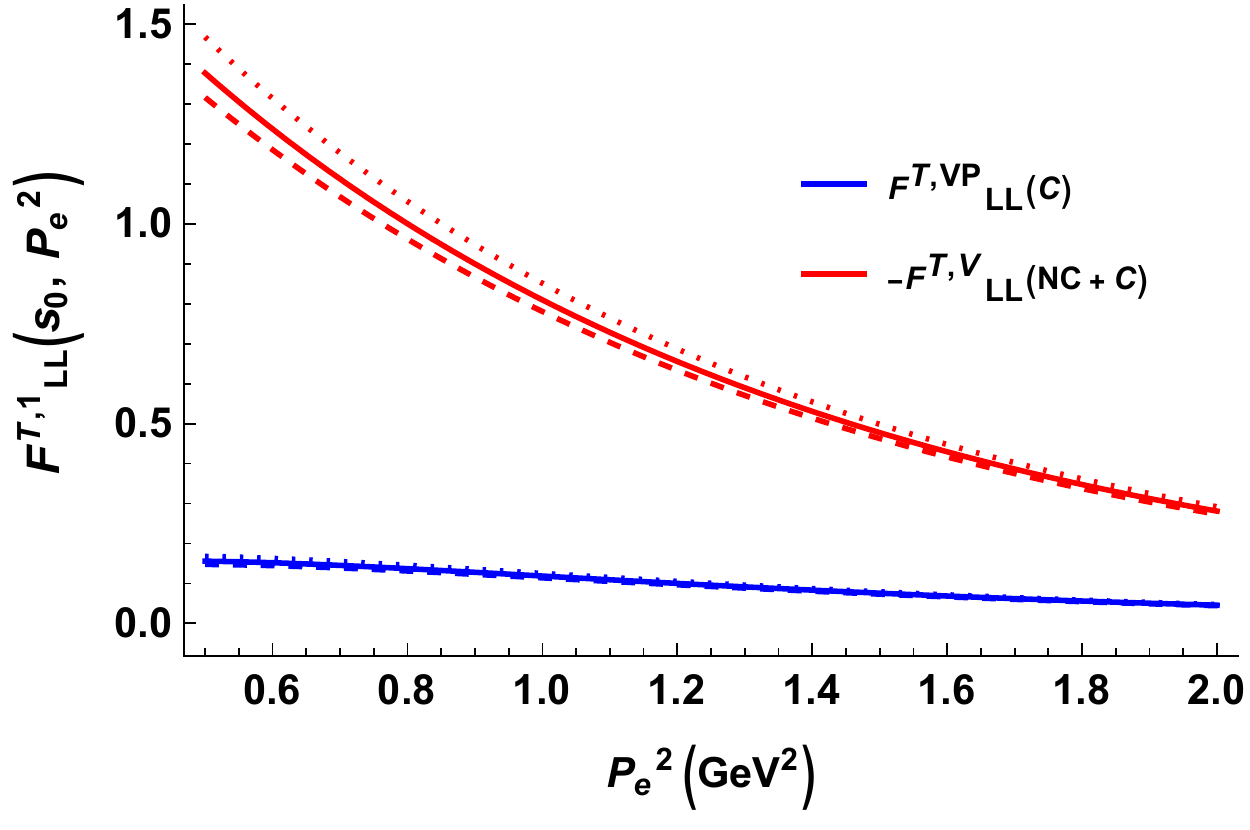}
   \end{subfigure}
   \begin{subfigure}{0.45\textwidth}
    \centering
    \includegraphics[width=0.8\linewidth]{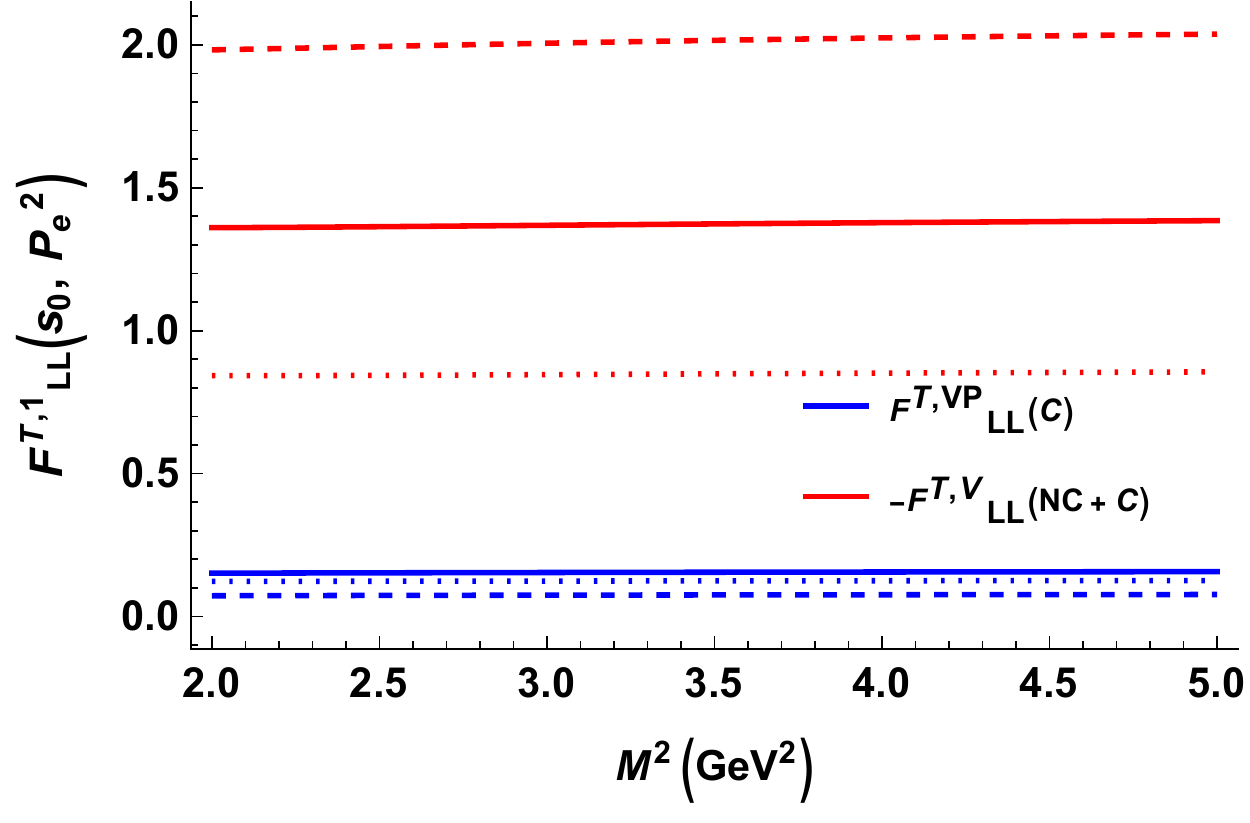}
   \end{subfigure}
    \caption{The FFs, $F_{LL}^{A,n}$ with $A\in\{S,V,T\}$ and $n\in\{0,1\}$ are extracted from different combinations of $F_{LL}^{A,r}$ with $r\in\{S,V,P,VP\}$ using the proton interpolation current $\chi_{IO}$. Left panel: We plot $F_{LL}^{A,n}$ vs $P_e^2$ for $s_0= (1.4\text{ GeV})^2$ (dashed), $s_0= (1.44\text{ GeV})^2$ (solid) and $s_0= (1.5\text{ GeV})^2$(dotted) with $M=2 \text{GeV}$ . Right panel: We plot $F_{LL}^{A,n}$ vs $M^2$ for $P_e^2= 0.1\text{ GeV}^2$ (dashed), $P_e^2= 0.5\text{ GeV}^2$ (solid) and $P_e^2= 1\text{ GeV}^2$ (dotted) with $s_0=(1.44 \text{GeV})^2$.}
    \label{LLtm1}
\end{figure}

\begin{figure}[h]
\centering
   \begin{subfigure}{0.45\textwidth}
    \centering
    \includegraphics[width=0.8\linewidth]{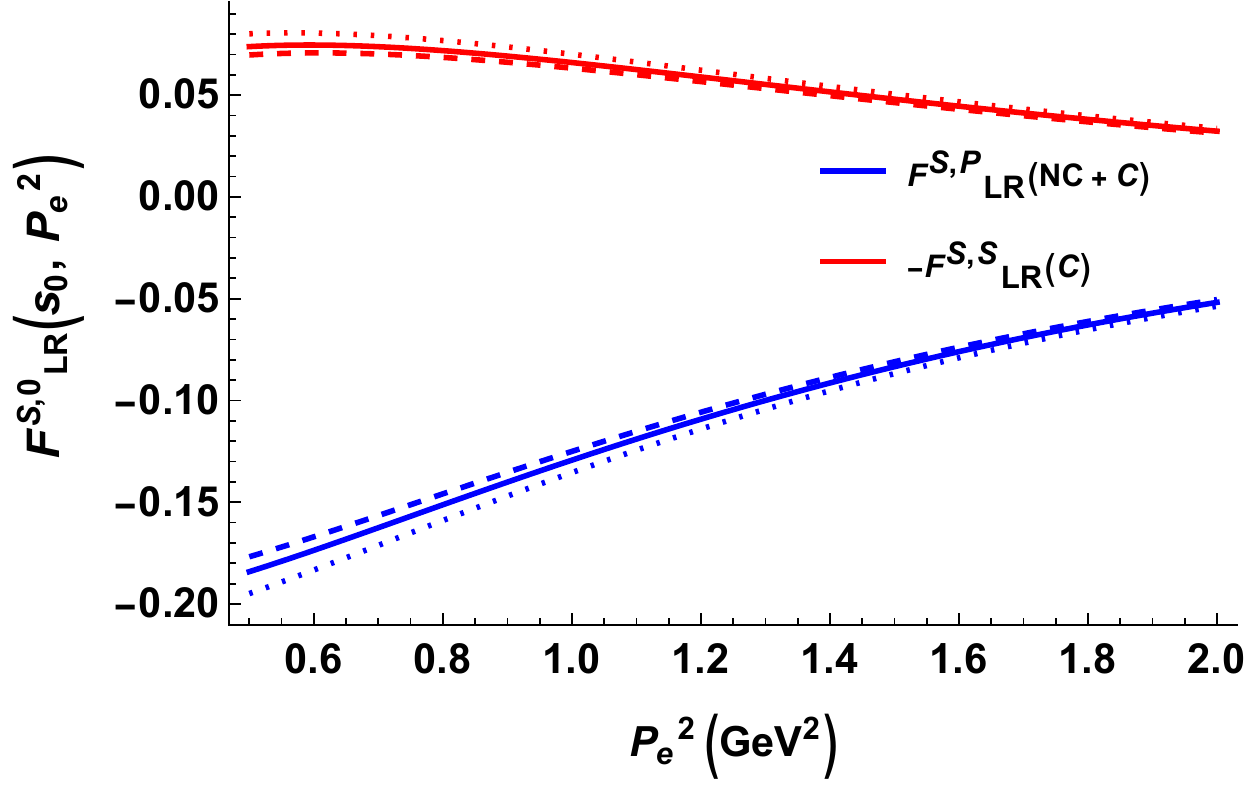}
   \end{subfigure}
   \begin{subfigure}{0.45\textwidth}
    \centering
    \includegraphics[width=0.8\linewidth]{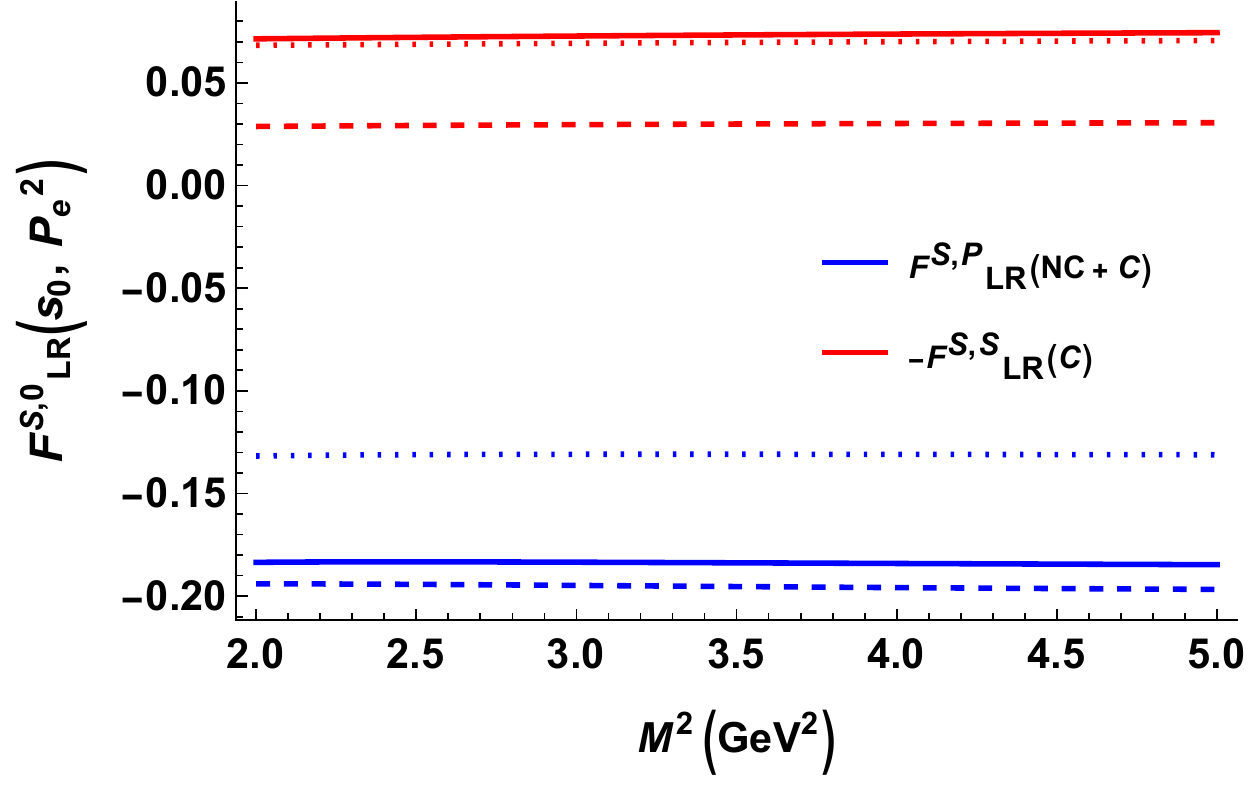}
   \end{subfigure}
   \begin{subfigure}{0.45\textwidth}
    \centering
    \includegraphics[width=0.8\linewidth]{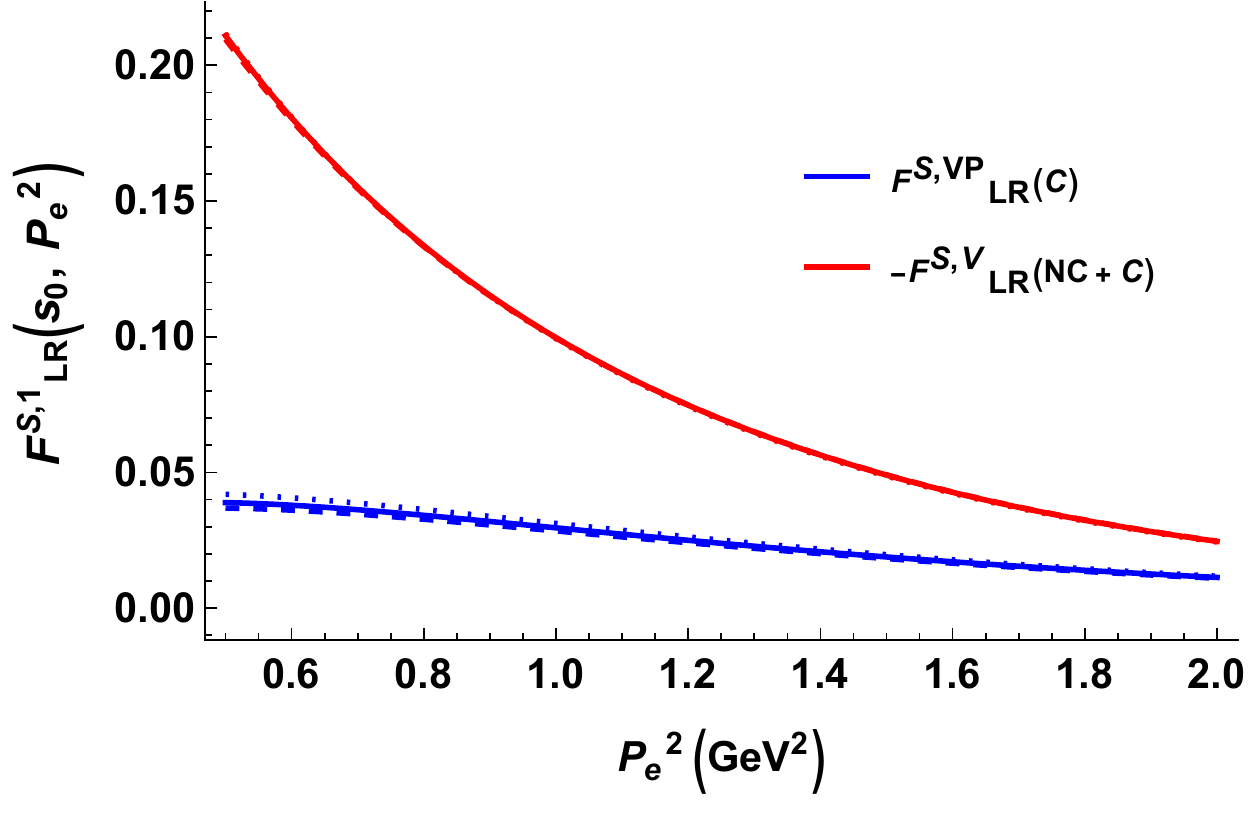}
   \end{subfigure}
   \begin{subfigure}{0.45\textwidth}
    \centering
    \includegraphics[width=0.8\linewidth]{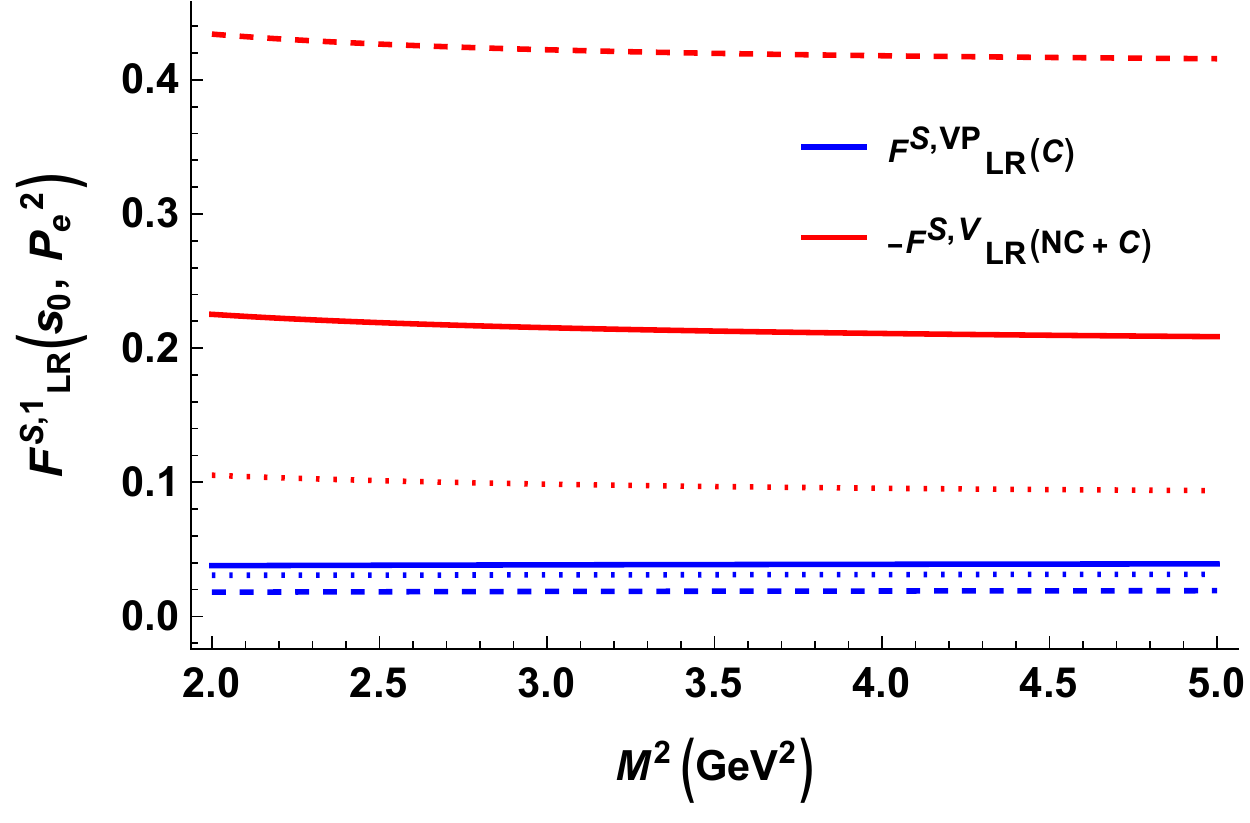}
   \end{subfigure}
   \begin{subfigure}{0.45\textwidth}
    \centering
    \includegraphics[width=0.8\linewidth]{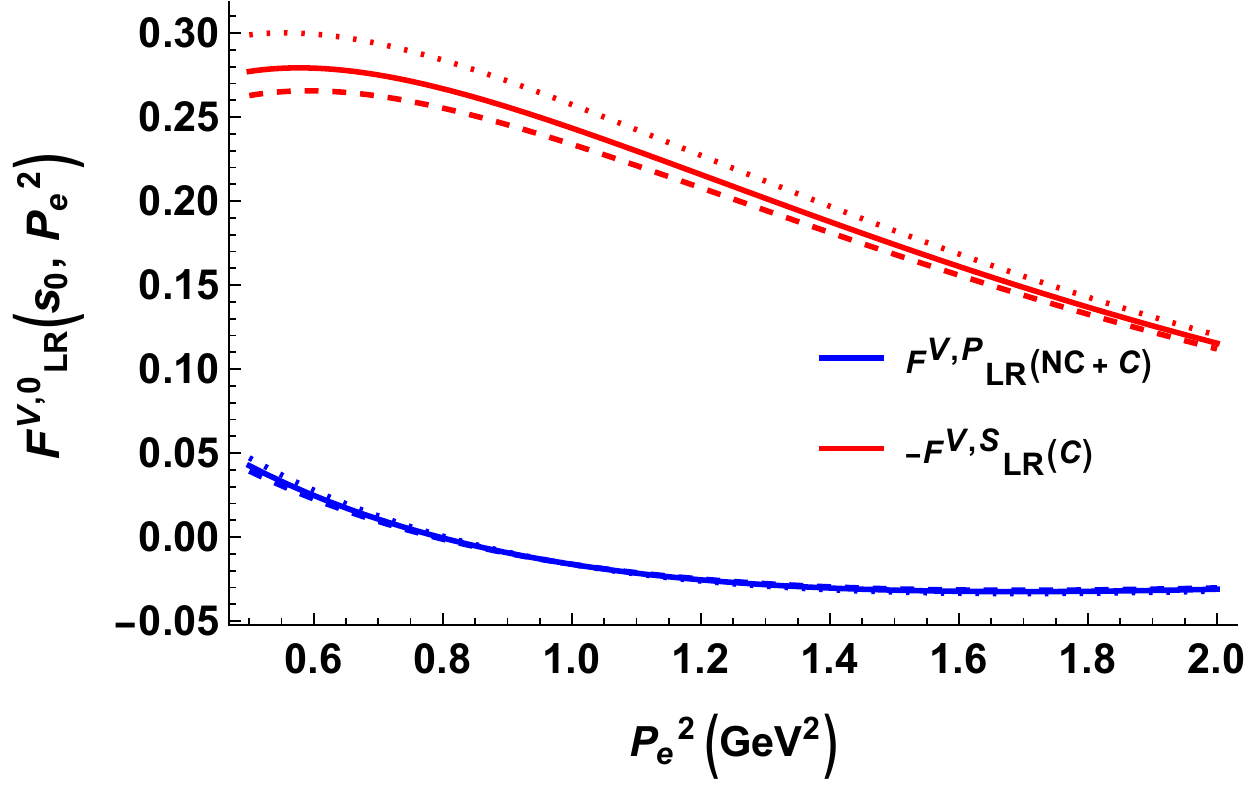}
   \end{subfigure}
   \begin{subfigure}{0.45\textwidth}
    \centering
    \includegraphics[width=0.8\linewidth]{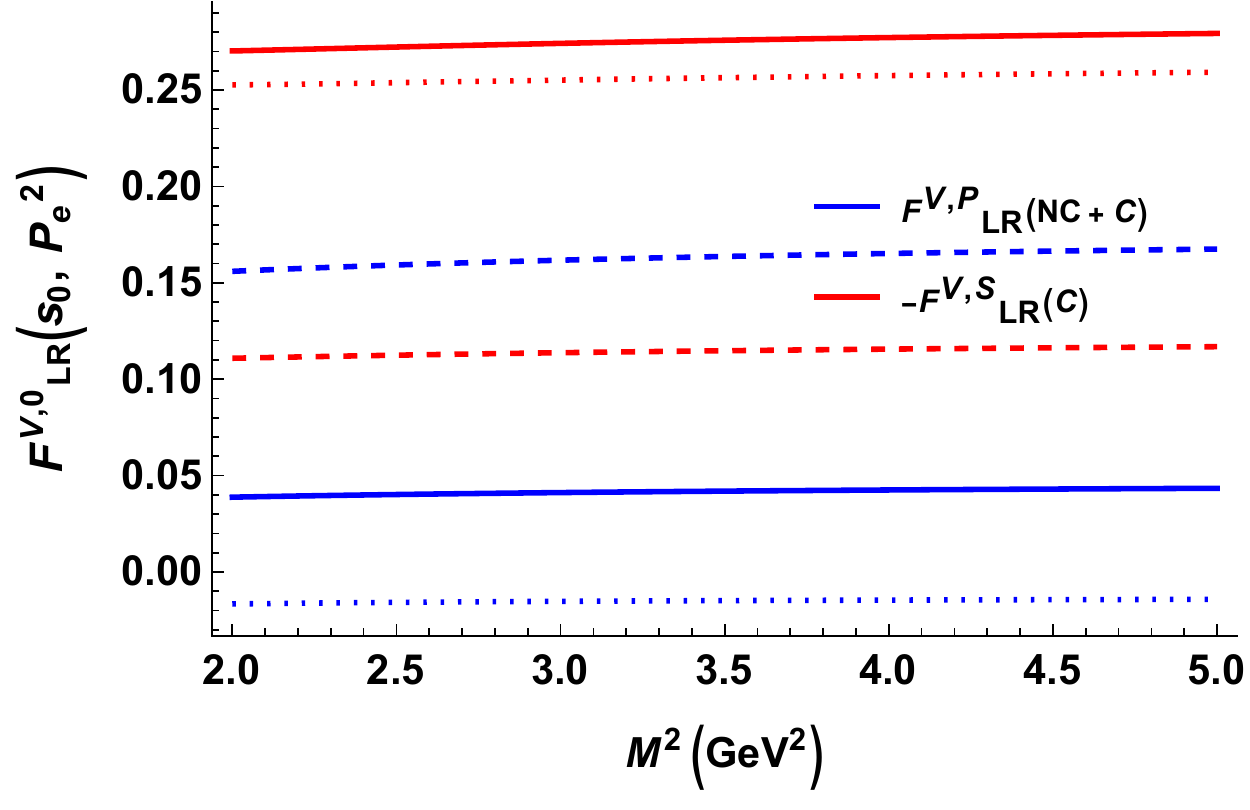}
   \end{subfigure}
   \begin{subfigure}{0.45\textwidth}
    \centering
    \includegraphics[width=0.8\linewidth]{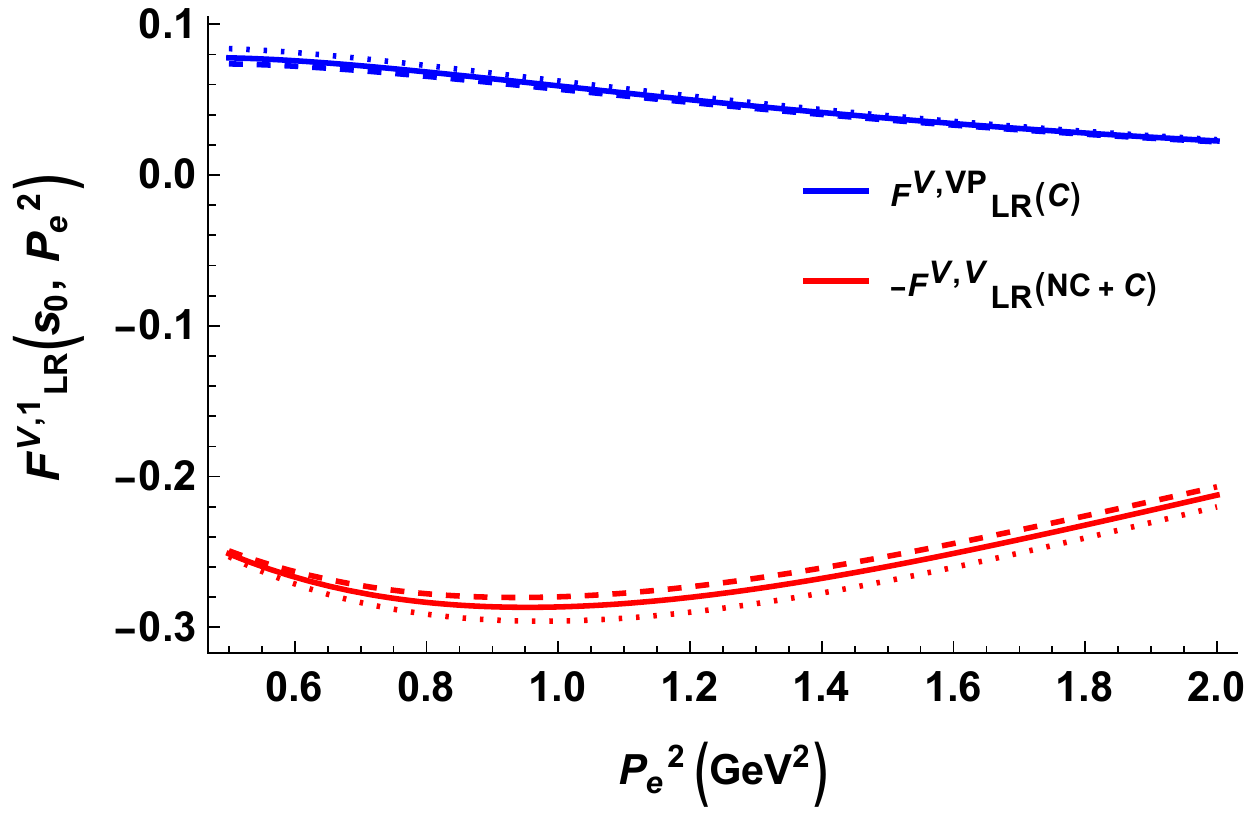}
   \end{subfigure}
   \begin{subfigure}{0.45\textwidth}
    \centering
    \includegraphics[width=0.8\linewidth]{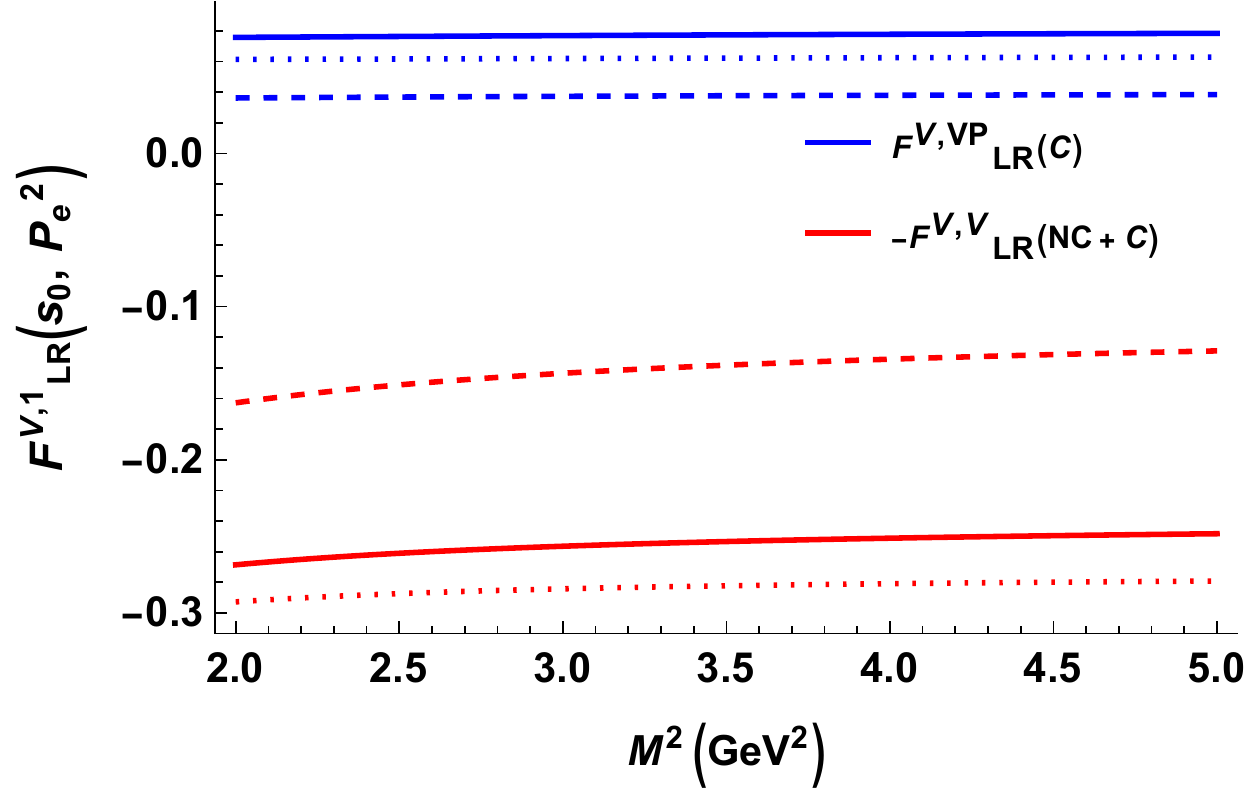}
   \end{subfigure}
    \caption{Same as Fig.(1) for $F_{LR}^{A,n}$ extracted from $F_{LR}^{A,r}$.}
    \label{LRtm1}
\end{figure}

The BNV process $D^0\rightarrow \bar p e^+$ involves 12 independent form factors (FFs). We have studied these FFs as a function of $P_e^2=-p_e^2$ and $M^2$ in the framework of LCSR using the distribution amplitudes of D-meson. As already discussed, the interpolation current for the proton state is not unique. To see the dependence of the FFs on the choice of interpolation current, we perform our numerical analysis for two different currents, $\chi_{IO}$ (see Eq.(\ref{chiio}) and Fig.(\ref{LLtm1} and \ref{LRtm1})) and $\chi_{LA}$ (see Eq.(\ref{chila}) and Fig.(\ref{LLt0} and \ref{LRt0})) and provide the analytical form for the general current in Appendix-B. $\chi_{IO}$ and $\chi_{LA}$ are the general choices of the proton interpolation current in LCSR and lattice QCD calculations, respectively. \\
In the sum rule, we have two independent parameters: the continuum threshold, $s_0$ and the Borel mass, $M$. The values of these are parameters are to be to be chosen such that the sum rule is saturated with the ground state contribution and the contribution coming from the continuum and the higher resonances should be well suppressed such that they do not contribute more than $30\%$ to the result. In order to do that, we choose $s_0=(1.44 \text{ GeV})^2$, the Roper resonance. This is the next resonance state after proton with the quantum numbers of the proton state. We have also shown the effect of varying $s_0$ on different FFs by taking three different values of $s_0$ (See Fig.(\ref{LLtm1}, \ref{LRtm1}, \ref{LLt0} and \ref{LRt0})). The Borel mass, $M$ should be chosen such that the form factors are stable with the variation in $M$ for a certain range of $M$ called the 'Borel window'. We have found the FFs to be almost stable for $M^2>2 \text{ GeV}^2$. We have shown the stability of the FFs in the Borel window for, $M^2 = (2-5) \text{ GeV}^2$ (See Fig.(\ref{LLtm1}, \ref{LRtm1}, \ref{LLt0} and \ref{LRt0})).

\begin{figure}[h]
\centering
   \begin{subfigure}{0.45\textwidth}
    \centering
    \includegraphics[width=0.8\linewidth]{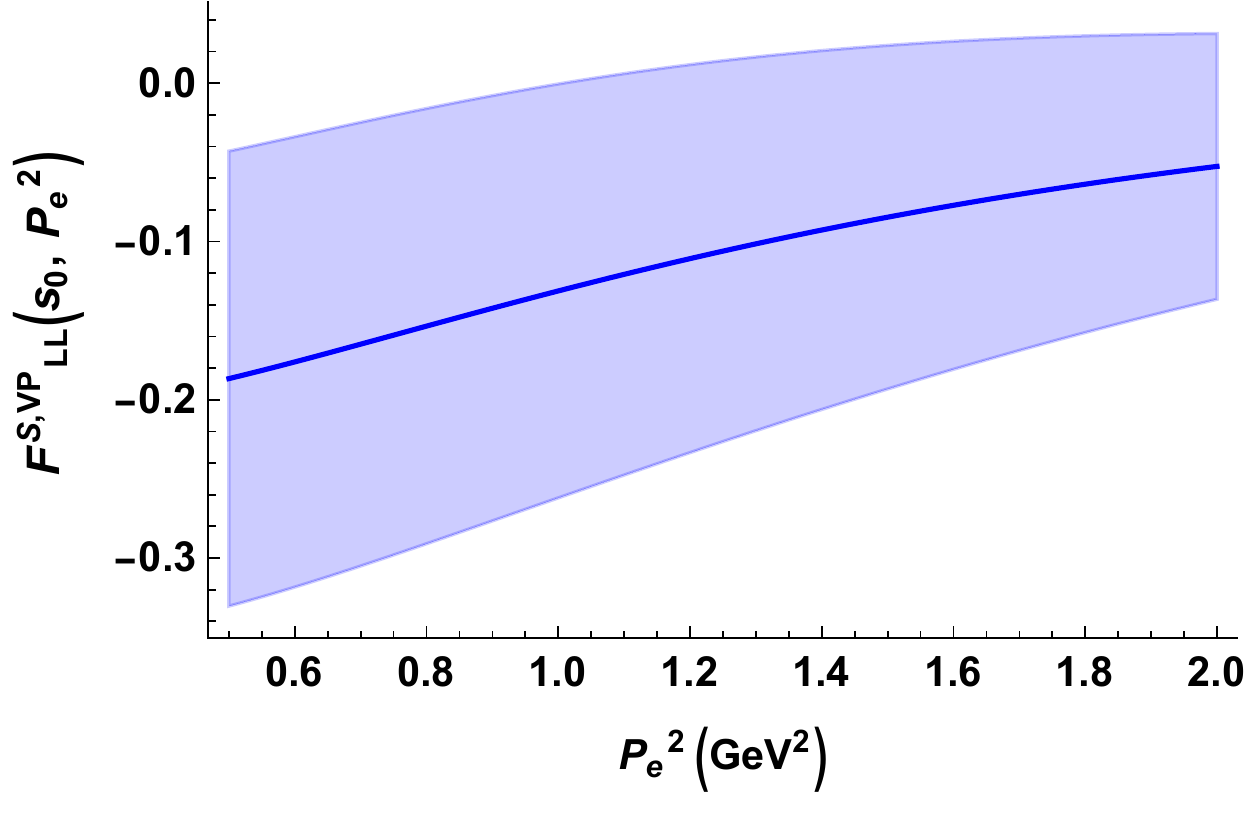}
   \end{subfigure}
   \begin{subfigure}{0.45\textwidth}
    \centering
    \includegraphics[width=0.8\linewidth]{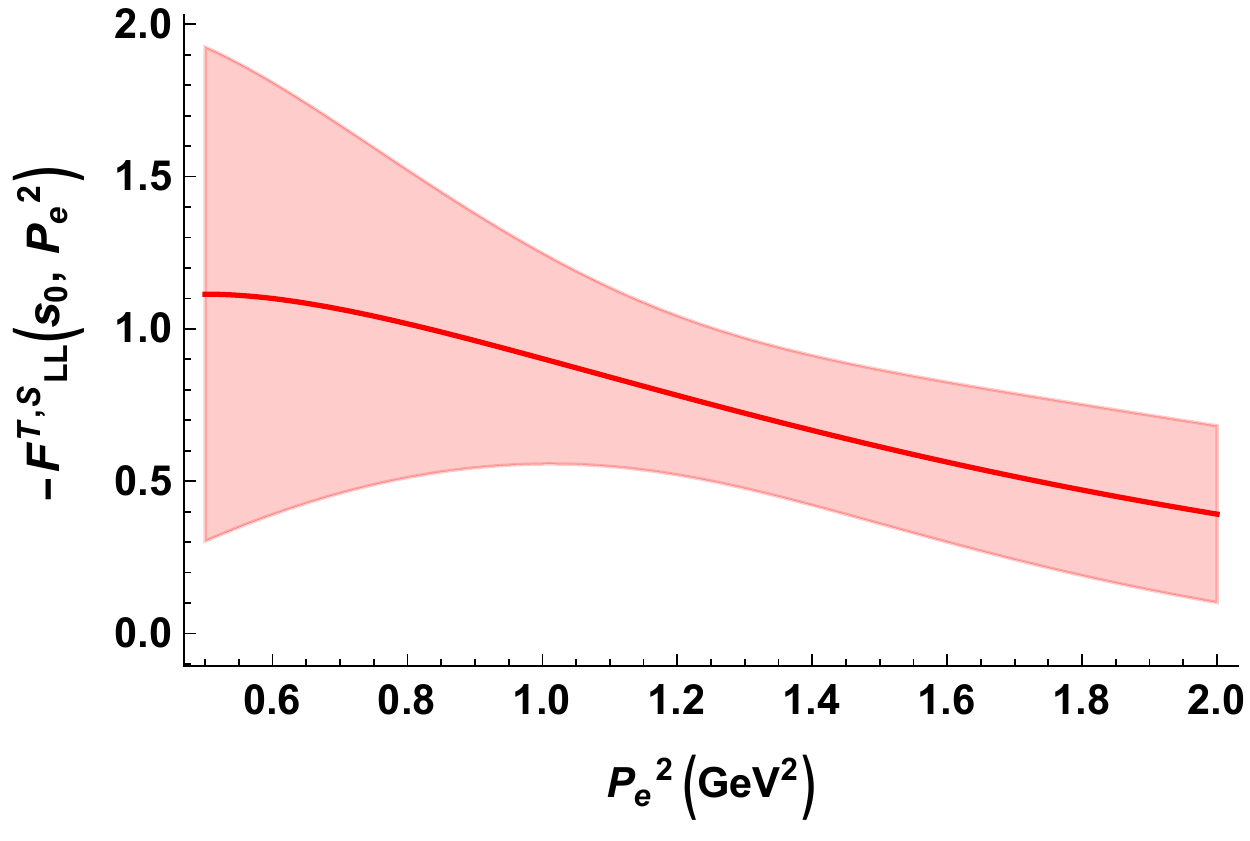}
   \end{subfigure}
   \begin{subfigure}{0.45\textwidth}
    \centering
    \includegraphics[width=0.8\linewidth]{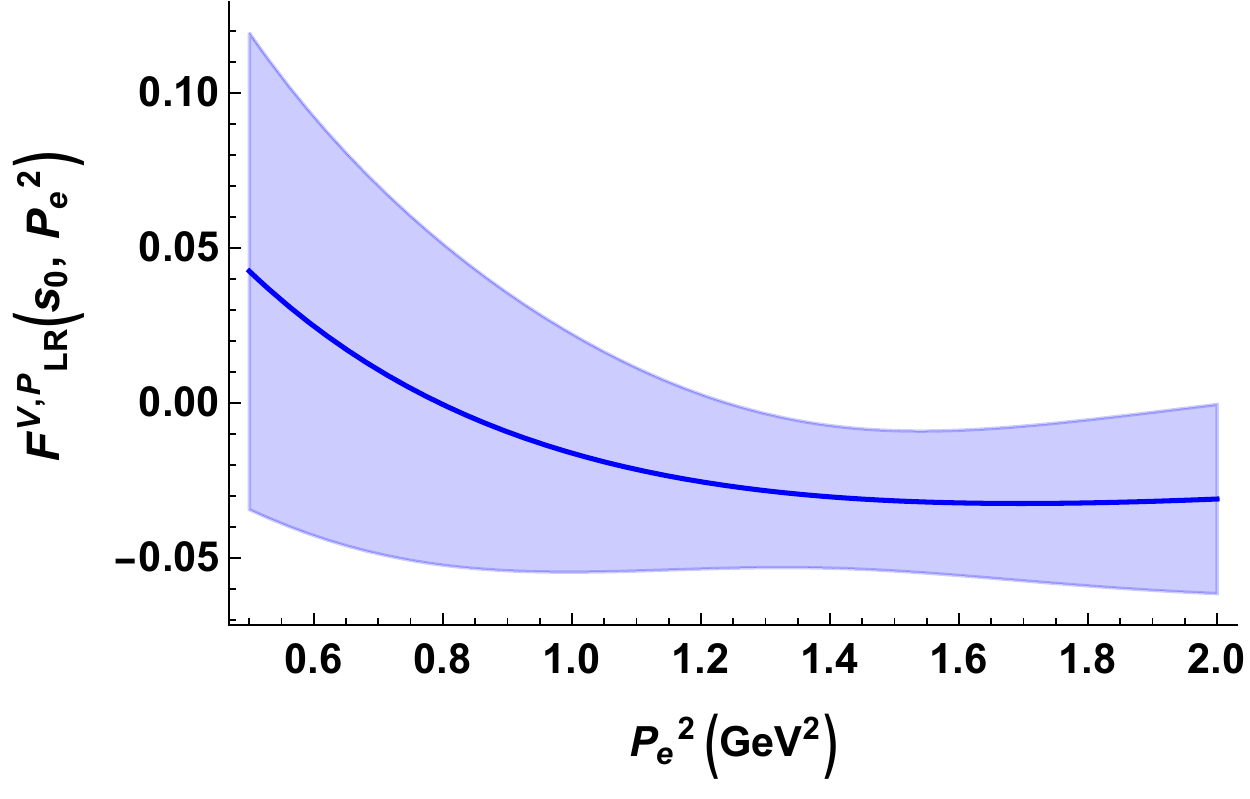}
   \end{subfigure}
   \begin{subfigure}{0.45\textwidth}
    \centering
    \includegraphics[width=0.8\linewidth]{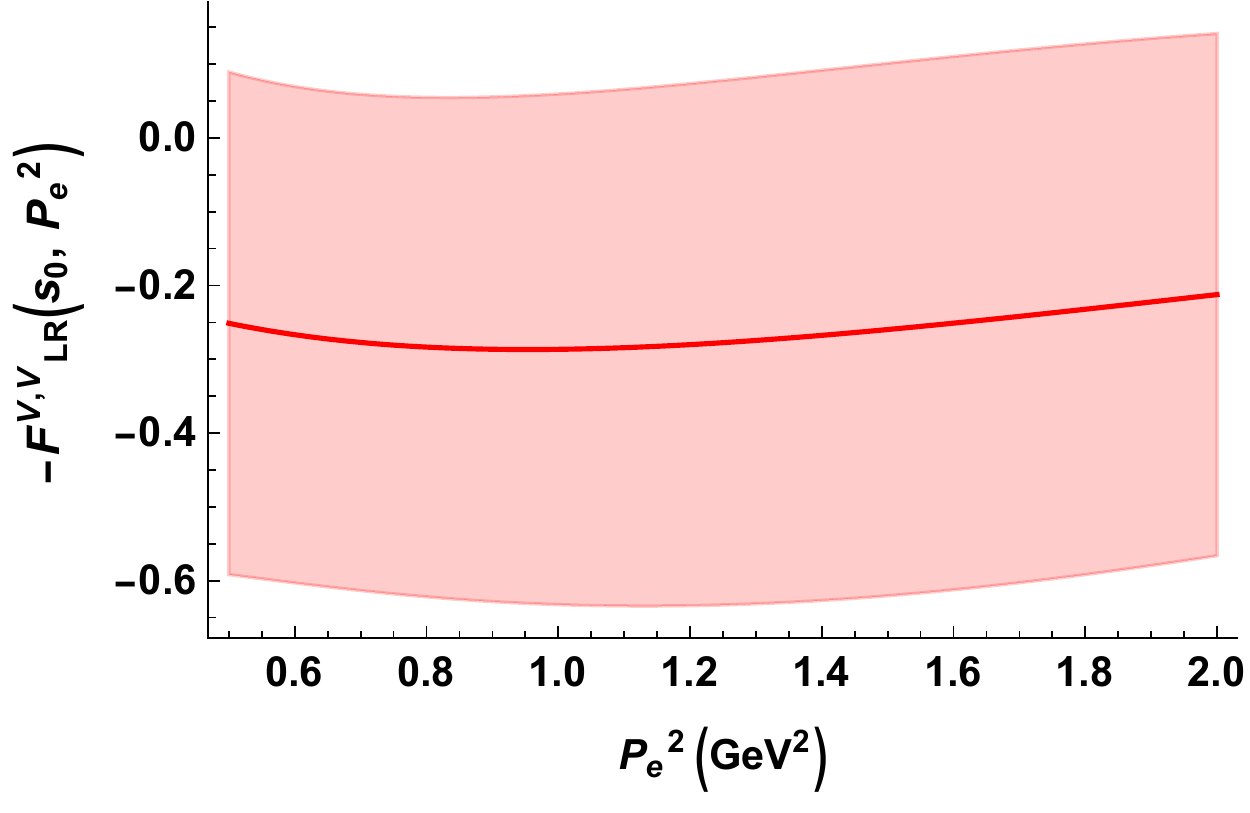}
   \end{subfigure}
    \caption{The representative graphs showing the variation of errors with $P_e^2$ for some $F_{\Gamma\Gamma'}^{A,r}$ functions calculated using the proton interpolation current, $\chi_{IO}$. The shaded regions represents the error band and the central line gives the calculated values of the $F_{\Gamma\Gamma'}^{A,r}$ functions}
    \label{etm1}
\end{figure}

\begin{figure}[h]
\centering
   \begin{subfigure}{0.45\textwidth}
    \centering
    \includegraphics[width=0.8\linewidth]{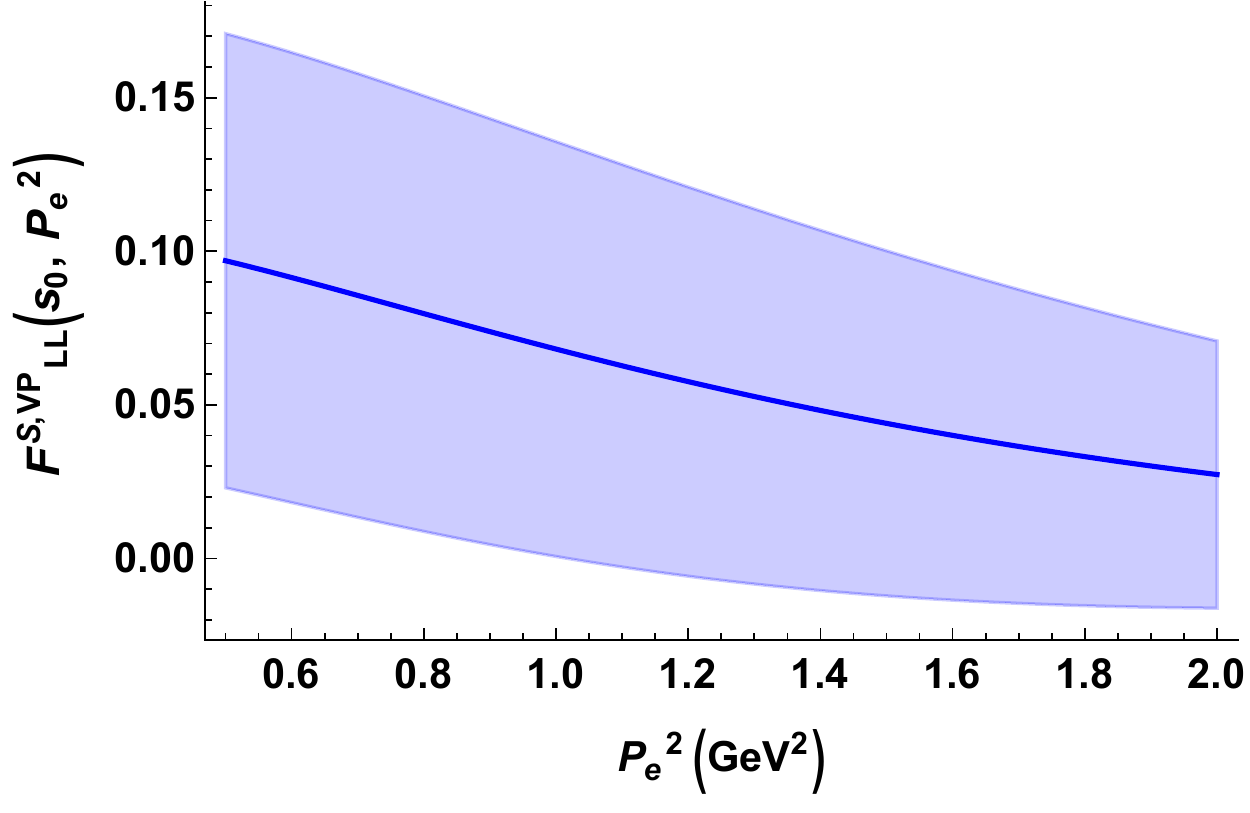}
   \end{subfigure}
   \begin{subfigure}{0.45\textwidth}
    \centering
    \includegraphics[width=0.8\linewidth]{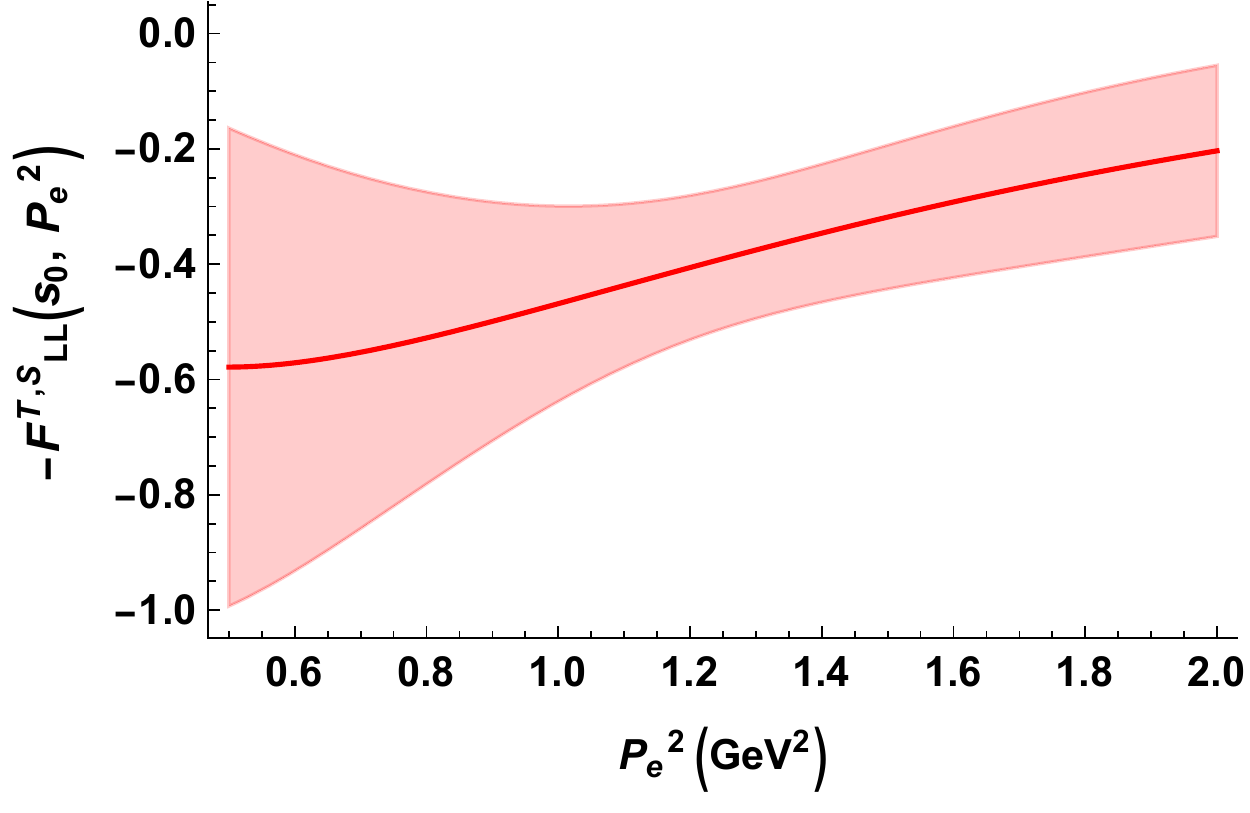}
   \end{subfigure}
   \begin{subfigure}{0.45\textwidth}
    \centering
    \includegraphics[width=0.8\linewidth]{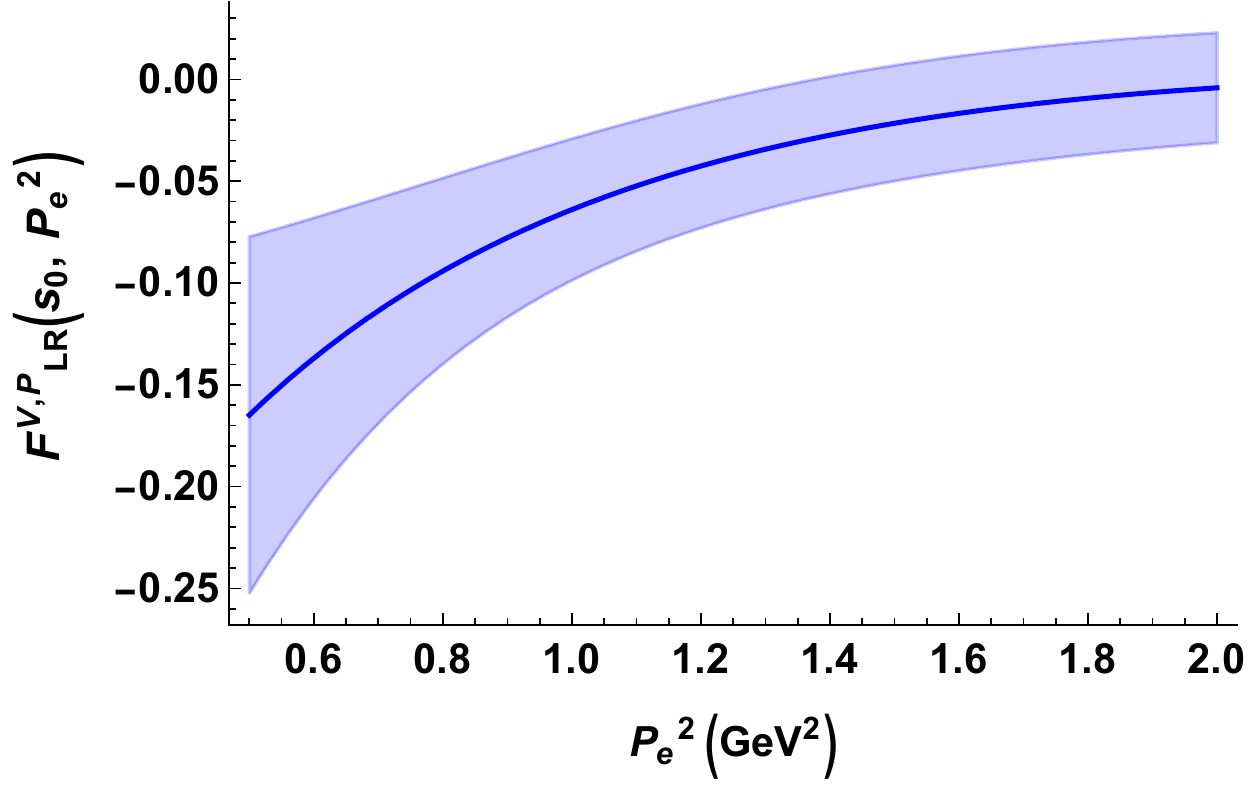}
   \end{subfigure}
   \begin{subfigure}{0.45\textwidth}
    \centering
    \includegraphics[width=0.8\linewidth]{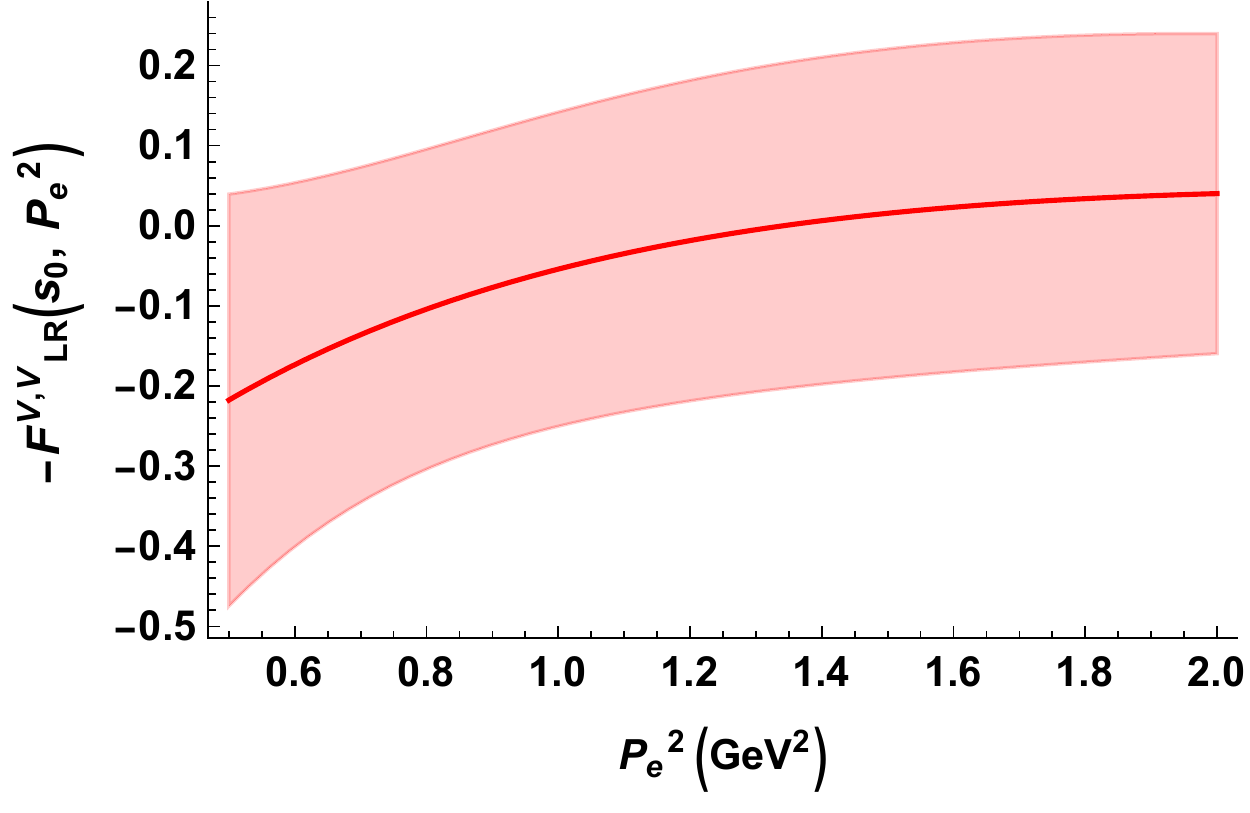}
   \end{subfigure}
    \caption{The representative graphs showing the variation of errors with $P_e^2$ for some $F_{\Gamma\Gamma'}^{A,r}$ functions calculated using the proton interpolation current, $\chi_{LA}$. The shaded regions represents the error band and the central line gives the calculated values of the $F_{\Gamma\Gamma'}^{A,r}$ functions}
    \label{et0}
\end{figure}

\begin{table}[h]
    \centering
    \begin{tabular}{|c|c|c||c|c|c|c}
    \hline
\multicolumn{3}{|c||}{Case-1: ($P_\Gamma=P_{\Gamma'}=P_L$)}&\multicolumn{3}{|c|}{Case-2: ($P_\Gamma=P_L,P_{\Gamma'}=P_R$)}\\
     \hline
     \hline
       Form Factor  & Extracted from & Value $\text{(GeV)}^2$&  Form Factor  & Extracted from & Value $\text{(GeV)}^2$\\
       \hline
         \multirow{2}{*}{$F_{LL}^{S,0}$}& $F_{LL}^{S,S}$&$0.211\pm0.471$ &\multirow{2}{*}{$F_{LR}^{S,0}$} &$F_{LR}^{S,S}$&$-0.106\pm0.075$\\ &$F_{LL}^{S,P}$&$0.041\pm0.036$& &$F_{LR}^{S,P}$&$0.074\pm0.143$ \\
         \hline
          \multirow{2}{*}{$F_{LL}^{S,1}$}& $F_{LL}^{S,V}$&$0.074\pm0.075$ &\multirow{2}{*}{$F_{LR}^{S,1}$} &$F_{LR}^{S,V}$&$0.211\pm0.471$\\ &$F_{LL}^{S,VP}$&$-0.187\pm0.144$& &$F_{LR}^{S,VP}$&$0.039\pm0.033$ \\
         \hline
          \multirow{2}{*}{$F_{LL}^{V,0}$}& $F_{LL}^{V,S}$&$0.360\pm0.467$ &\multirow{2}{*}{$F_{LR}^{V,0}$} &$F_{LR}^{V,S}$&$0.277\pm0.227$\\ &$F_{LL}^{V,P}$&$0.099\pm0.036$& &$F_{LR}^{V,P}$&$0.043\pm0.077$ \\
         \hline
          \multirow{2}{*}{$F_{LL}^{V,1}$}& $F_{LL}^{V,V}$&$0.271\pm0.097$ &\multirow{2}{*}{$F_{LR}^{V,1}$} &$F_{LR}^{V,V}$&$-0.251\pm0.341$\\ &$F_{LL}^{V,VP}$&$-0.067\pm0.154$& &$F_{LR}^{V,VP}$&$0.078\pm0.067$ \\
         \hline
          \multirow{2}{*}{$F_{LL}^{T,0}$}&$F_{LL}^{T,S}$&$1.114\pm0.812$ &\multirow{2}{*}{$F_{LR}^{T,0}$} &$F_{LR}^{T,S}$&0\\ &$F_{LL}^{T,P}$&$0.550\pm0.256$& &$F_{LR}^{T,P}$&0 \\
         \hline
          \multirow{2}{*}{$F_{LL}^{T,1}$}&$F_{LL}^{T,V}$&$1.378\pm0.629$ &\multirow{2}{*}{$F_{LR}^{T,1}$} &$F_{LR}^{T,V}$&0\\ &$F_{LL}^{T,VP}$&$0.156\pm0.133$& &$F_{LR}^{T,VP}$&0 \\
         \hline
         
    \end{tabular}
    \caption{Tabulation of all the 12 independent FFs at $P_e^2= 0.5 \text{ GeV}^2$ for $s_0=(1.44\text{ GeV} )^2$ and $M=2 \text{ GeV}$ calculated using the proton interpolation current $\chi_{IO}$. The errors are associated with the errors in the parameters used for the numerical analysis.}
    \label{LRNAvaluetm1}
\end{table}
\begin{table}[h]
    \centering
    \begin{tabular}{|c|c|c||c|c|c|c}
    \hline
\multicolumn{3}{|c||}{Case-1: ($P_\Gamma=P_{\Gamma'}=P_L$)}&\multicolumn{3}{|c|}{Case-2: ($P_\Gamma=P_L,P_{\Gamma'}=P_R$)}\\
     \hline
     \hline
       Form Factor  & Extracted from & Value $\text{(GeV)}^2$&  Form Factor  & Extracted from & Value $\text{(GeV)}^2$\\
       \hline
         \multirow{2}{*}{$F_{LL}^{S,0}$}& $F_{LL}^{S,S}$&$-0.110\pm0.244$ &\multirow{2}{*}{$F_{LR}^{S,0}$} &$F_{LR}^{S,S}$&$0.183\pm0.363$\\ &$F_{LL}^{S,P}$&$-0.135\pm0.109$& &$F_{LR}^{S,P}$&$0.096\pm0.074$ \\
         \hline
          \multirow{2}{*}{$F_{LL}^{S,1}$}& $F_{LL}^{S,V}$&$0.183\pm0.363$ &\multirow{2}{*}{$F_{LR}^{S,1}$} &$F_{LR}^{S,V}$&$-0.110\pm0.244$\\ &$F_{LL}^{S,VP}$&$0.097\pm0.074$& &$F_{LR}^{S,VP}$&$-0.133\pm0.109$ \\
         \hline
          \multirow{2}{*}{$F_{LL}^{V,0}$}& $F_{LL}^{V,S}$&$-0.302\pm0.268$ &\multirow{2}{*}{$F_{LR}^{V,0}$} &$F_{LR}^{V,S}$&$-0.345\pm0.270$\\ &$F_{LL}^{V,P}$&$-0.062\pm0.026$& &$F_{LR}^{V,P}$&$-0.165\pm0.088$ \\
         \hline
          \multirow{2}{*}{$F_{LL}^{V,1}$}& $F_{LL}^{V,V}$&$-0.204\pm0.072$ &\multirow{2}{*}{$F_{LR}^{V,1}$} &$F_{LR}^{V,V}$&$0.218\pm0.257$\\ &$F_{LL}^{V,VP}$&$-0.046\pm0.125$& &$F_{LR}^{V,VP}$&$0.081\pm0.068$ \\
         \hline
          \multirow{2}{*}{$F_{LL}^{T,0}$}&$F_{LL}^{T,S}$&$-0.578\pm0.414$ &\multirow{2}{*}{$F_{LR}^{T,0}$} &$F_{LR}^{T,S}$&0\\ &$F_{LL}^{T,P}$&$-0.018\pm0.152$& &$F_{LR}^{T,P}$&0 \\
         \hline
          \multirow{2}{*}{$F_{LL}^{T,1}$}&$F_{LL}^{T,V}$&$-0.081\pm0.405$ &\multirow{2}{*}{$F_{LR}^{T,1}$} &$F_{LR}^{T,V}$&0\\ &$F_{LL}^{T,VP}$&$0.034\pm0.068$& &$F_{LR}^{T,VP}$&0 \\
         \hline
         
    \end{tabular}
    \caption{Tabulation of all the 12 independent FFs at $P_e^2= 0.5 \text{ GeV}^2$ for $s_0=(1.44\text{ GeV} )^2$ and $M=2 \text{ GeV}$ calculated using the proton interpolation current $\chi_{LA}$. The errors are associated with the errors in the parameters used for the numerical analysis.} 
    \label{LRNAvaluet0}
\end{table}
Each form factor can be calculated from two $F_{\Gamma\Gamma'}^{A,r}$ as given in Eq.(3.7). It has been found that some of these $F_{\Gamma\Gamma'}^{A,r}$ get contribution only from the condensate. Due to the presence of only condensate contributions in some $F_{\Gamma\Gamma'}^{A,r}$, we found the difference in the extraction of the FFs using different combinations of $F_{\Gamma\Gamma'}^{A,r}$. We tabulate these form factors in Table-\ref{LRNAvaluetm1} and Table-\ref{LRNAvaluet0} for $\chi_{IO}$ and $\chi_{LA}$, respectively at $P_e^2=0.5 \text{ GeV}^2$. In Figs.(\ref{LLtm1},\ref{LRtm1},\ref{LLt0} and \ref{LRt0}), we have labelled these different combinations with (C) and (NC+C) for having only the condensate contribution and having condensate as well as non-condensate contributions, respectively. The two FFs, $F_{LR}^{T,0}$ and $F_{LR}^{T,1}$ are found to be explicitly zero. We have also calculated the errors in the FFs associated with the uncertainties in the values of the parameters used for the numerical analysis. We have found that the errors are very large (even upto $200\%$) in some cases as can be seen from Table-\ref{LRNAvaluetm1} and Table-\ref{LRNAvaluet0}. In Fig.(\ref{etm1}) and Fig.(\ref{et0}), we show some representative graphs showing the variation of error in $F_{\Gamma\Gamma'}^{A,r}$ with $P_e^2$ for the  proton interpolations currents, $\chi_{IO}$ and $\chi_{LA}$, respectively. The errors are found to be dominated by the uncertainty in $w_0$ which is a model input parameter in the LCDAs of D-meson. 
\begin{figure}[h]
\centering
   \begin{subfigure}{0.45\textwidth}
    \centering
    \includegraphics[width=0.8\linewidth]{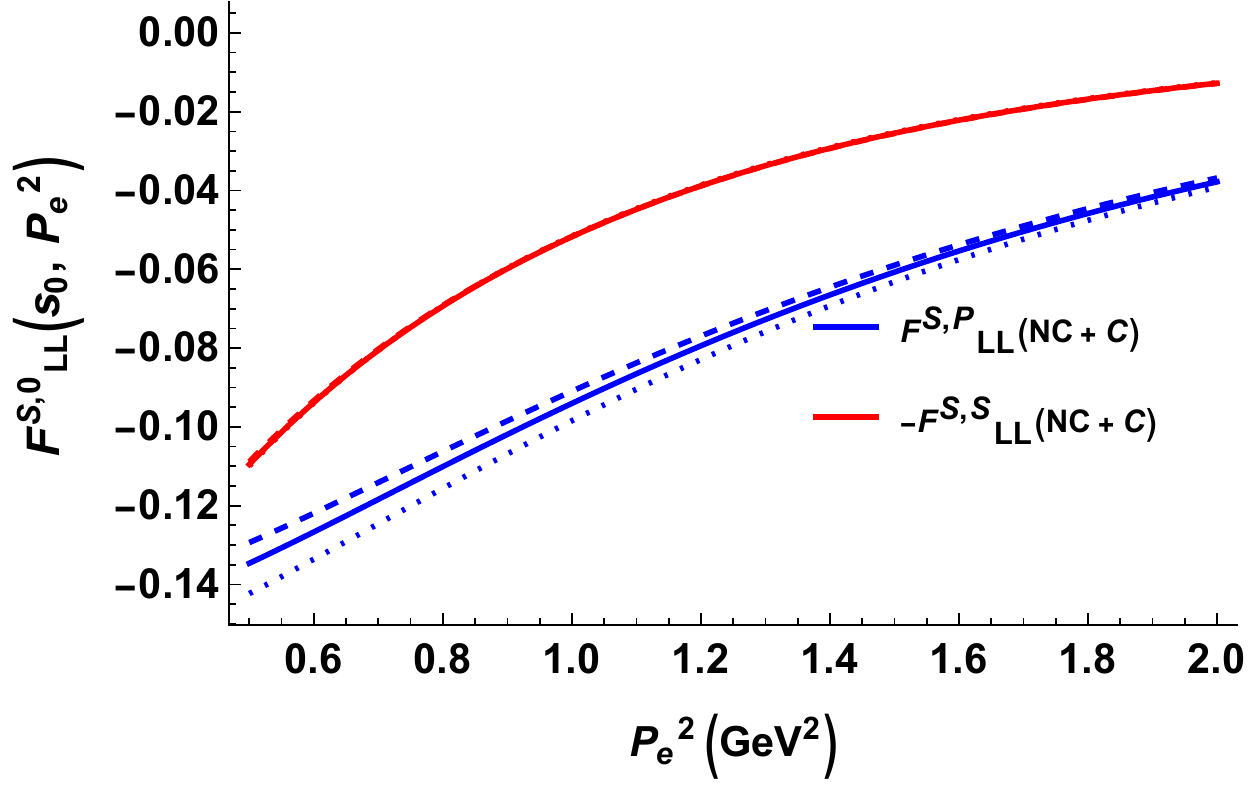}
   \end{subfigure}
   \begin{subfigure}{0.45\textwidth}
    \centering
    \includegraphics[width=0.8\linewidth]{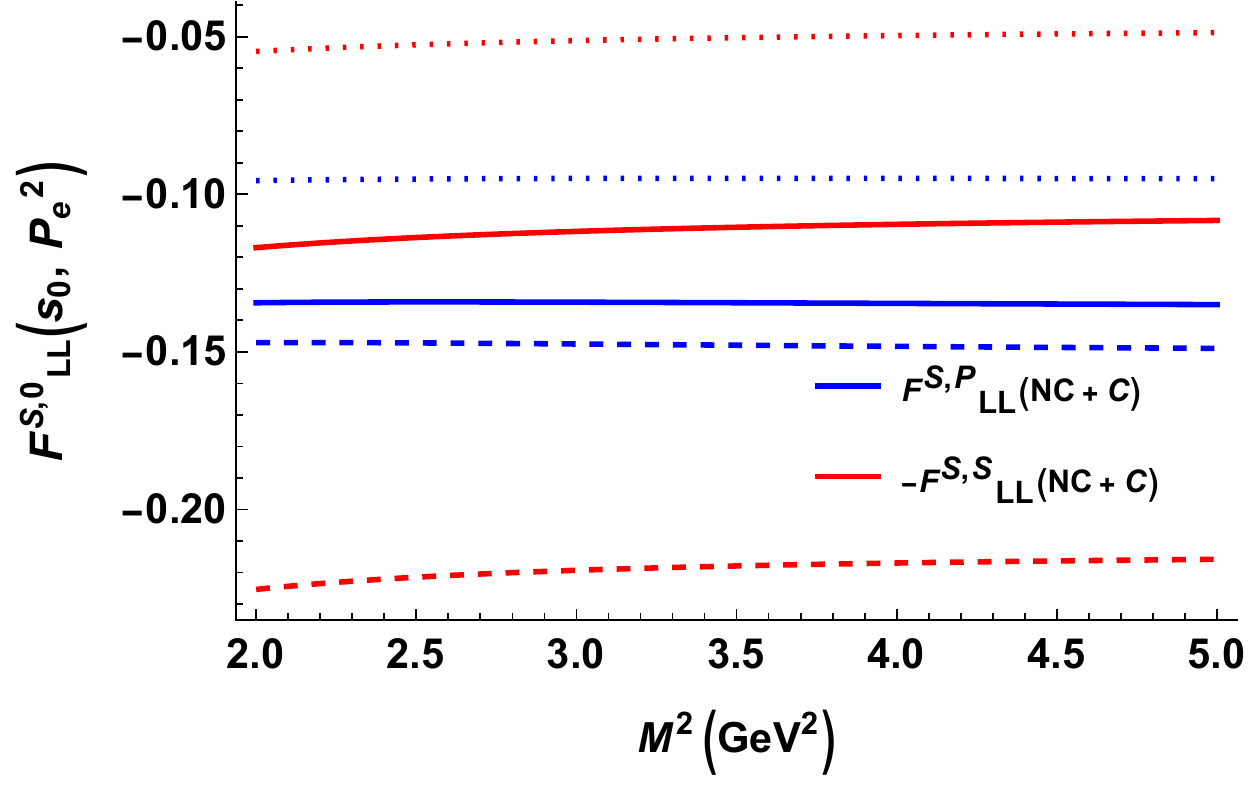}
   \end{subfigure}
   \begin{subfigure}{0.45\textwidth}
    \centering
    \includegraphics[width=0.8\linewidth]{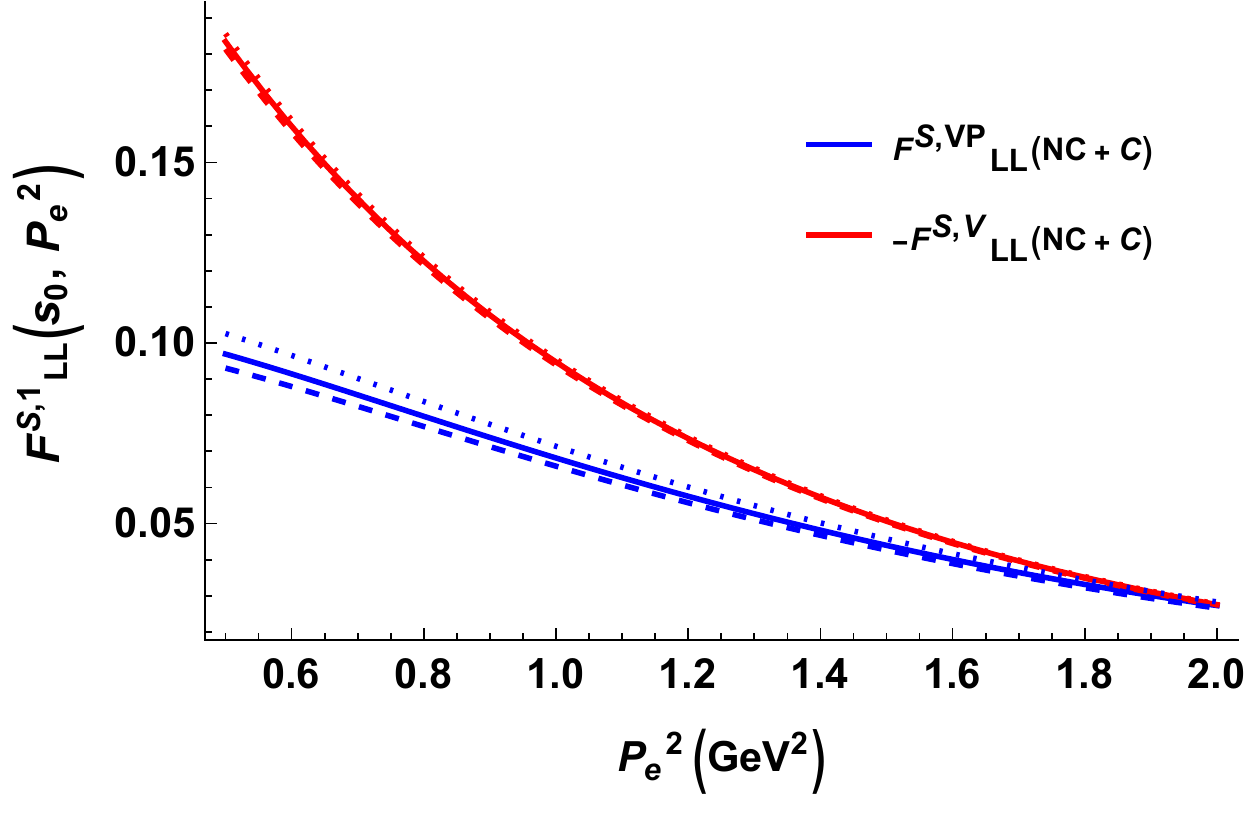}
   \end{subfigure}
   \begin{subfigure}{0.45\textwidth}
    \centering
    \includegraphics[width=0.8\linewidth]{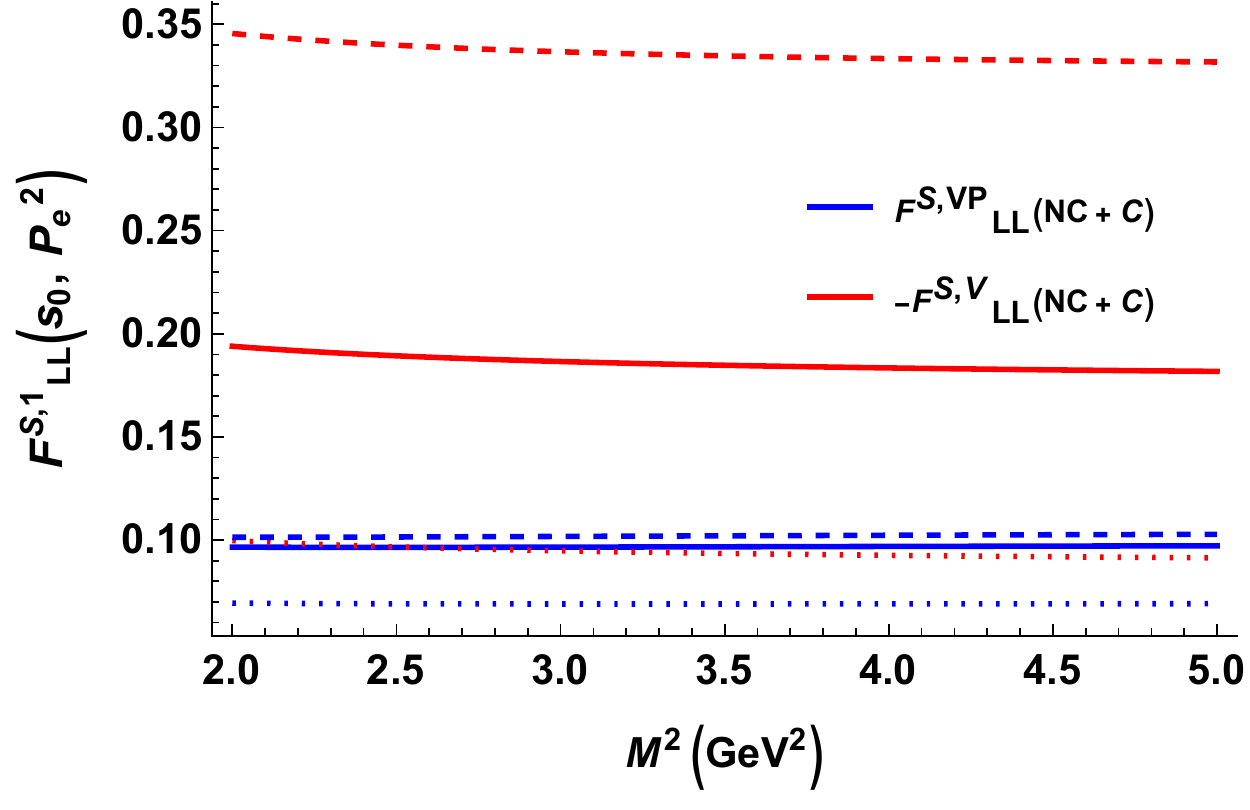}
   \end{subfigure}
   \begin{subfigure}{0.45\textwidth}
    \centering
    \includegraphics[width=0.8\linewidth]{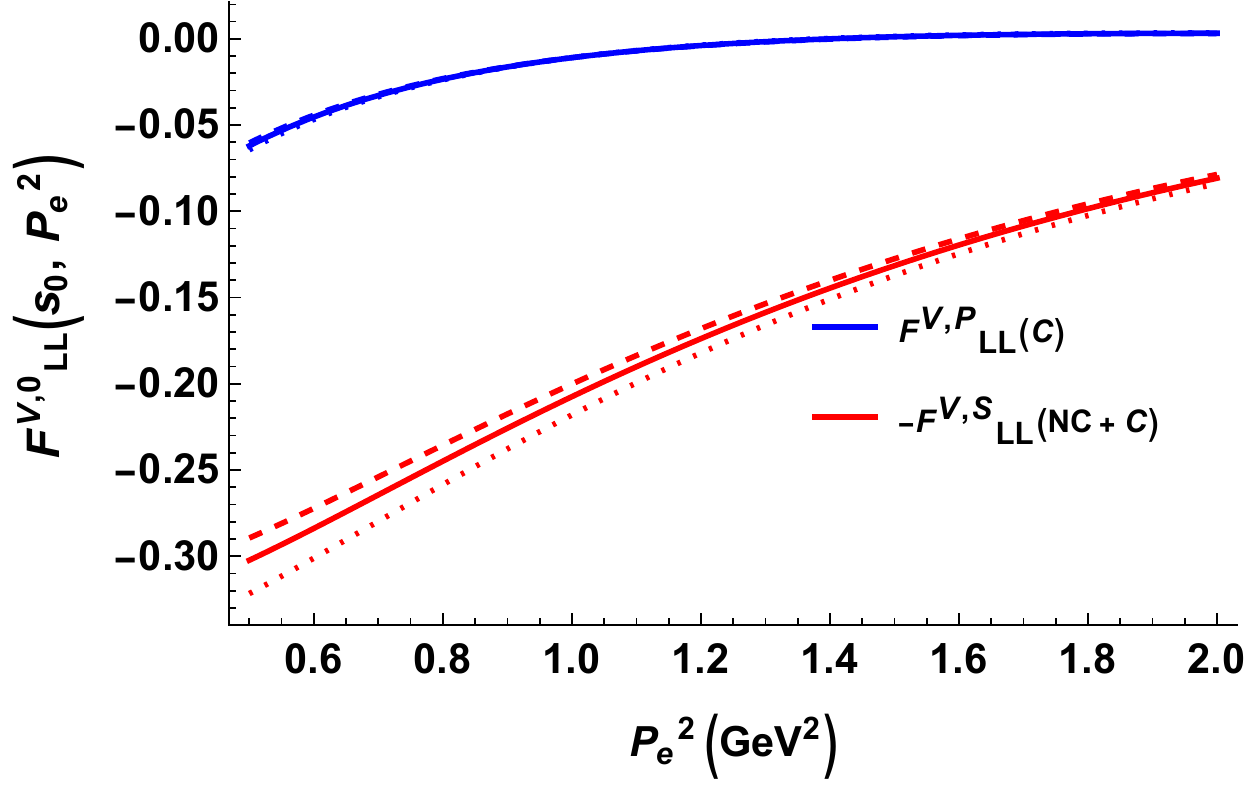}
   \end{subfigure}
   \begin{subfigure}{0.45\textwidth}
    \centering
    \includegraphics[width=0.8\linewidth]{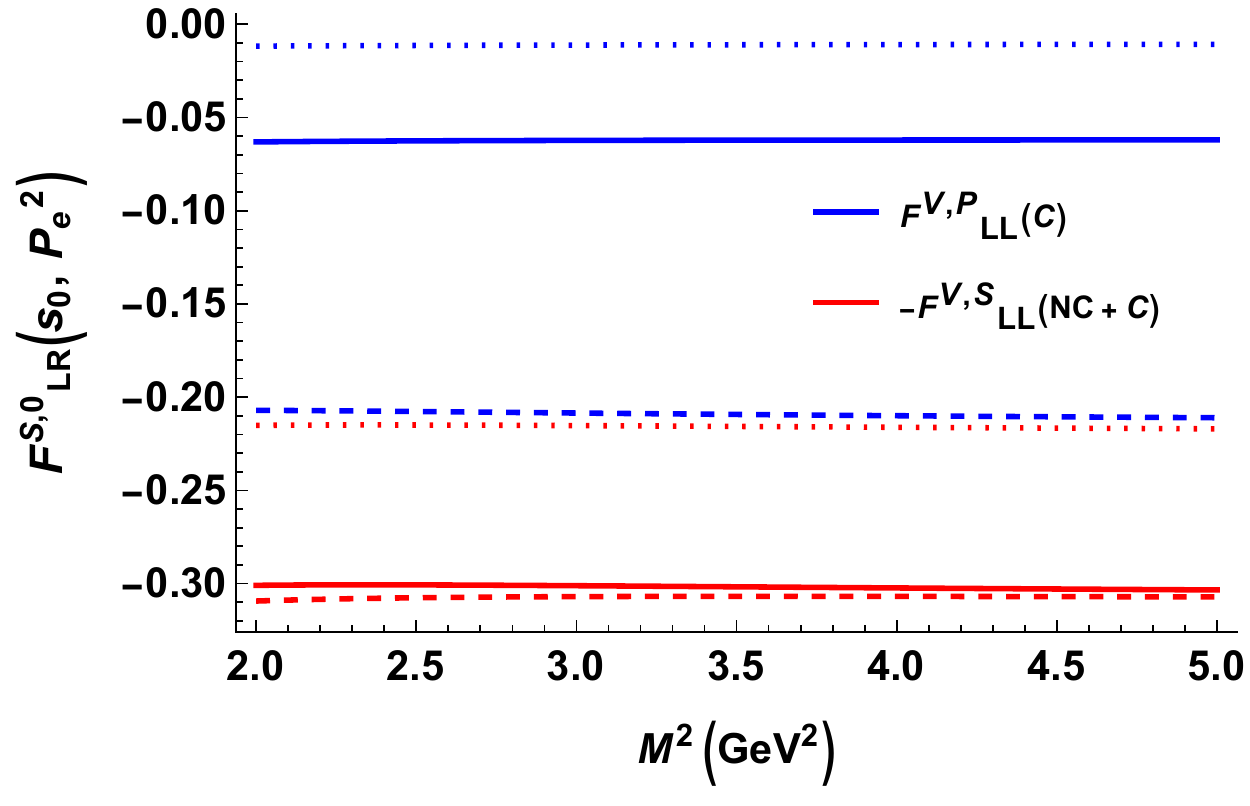}
   \end{subfigure}
   \begin{subfigure}{0.45\textwidth}
    \centering
    \includegraphics[width=0.8\linewidth]{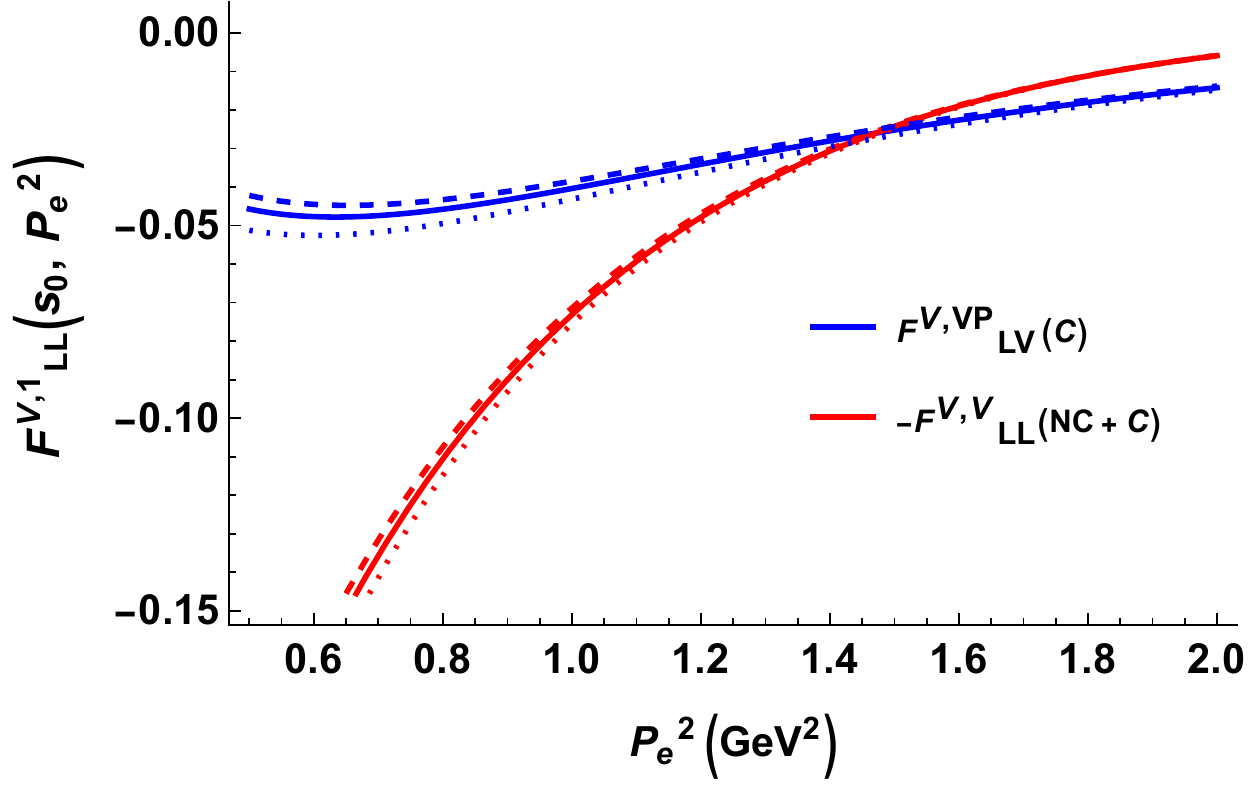}
   \end{subfigure}
   \begin{subfigure}{0.45\textwidth}
    \centering
    \includegraphics[width=0.8\linewidth]{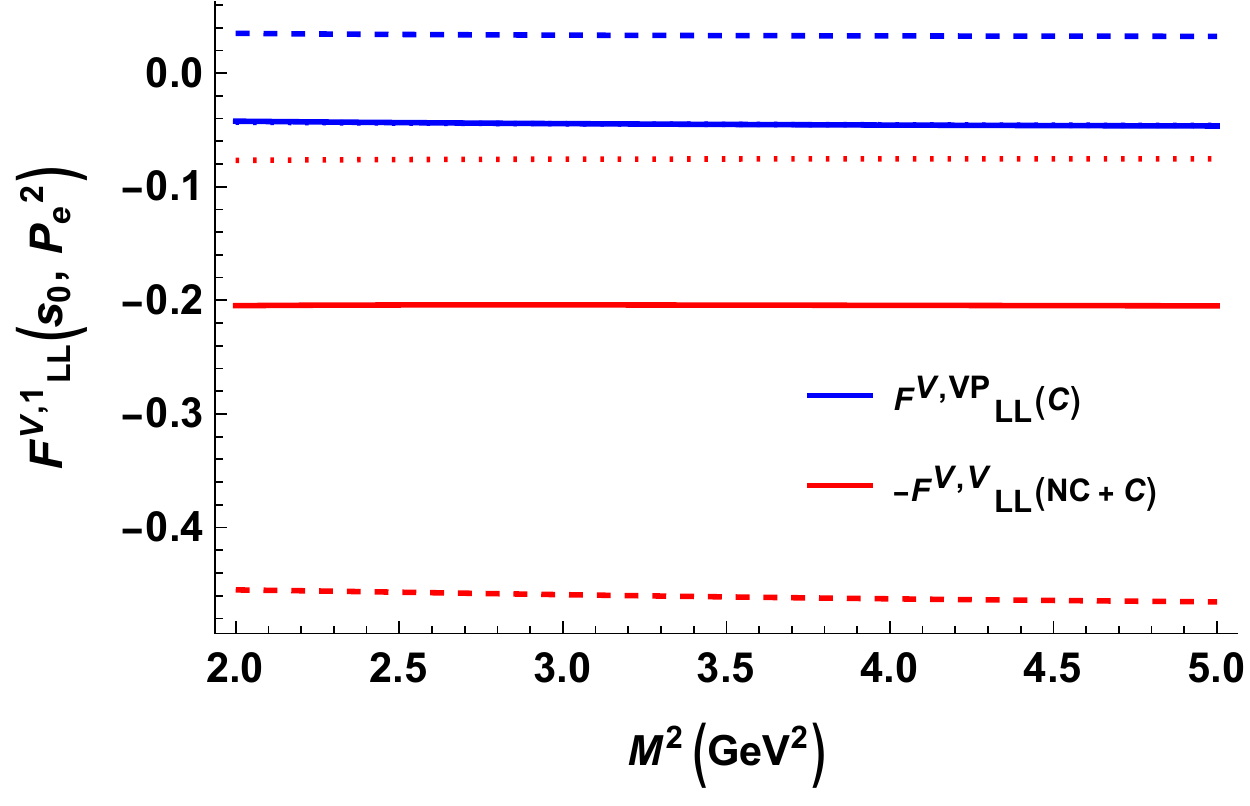}
   \end{subfigure}
   \begin{subfigure}{0.45\textwidth}
    \centering
    \includegraphics[width=0.8\linewidth]{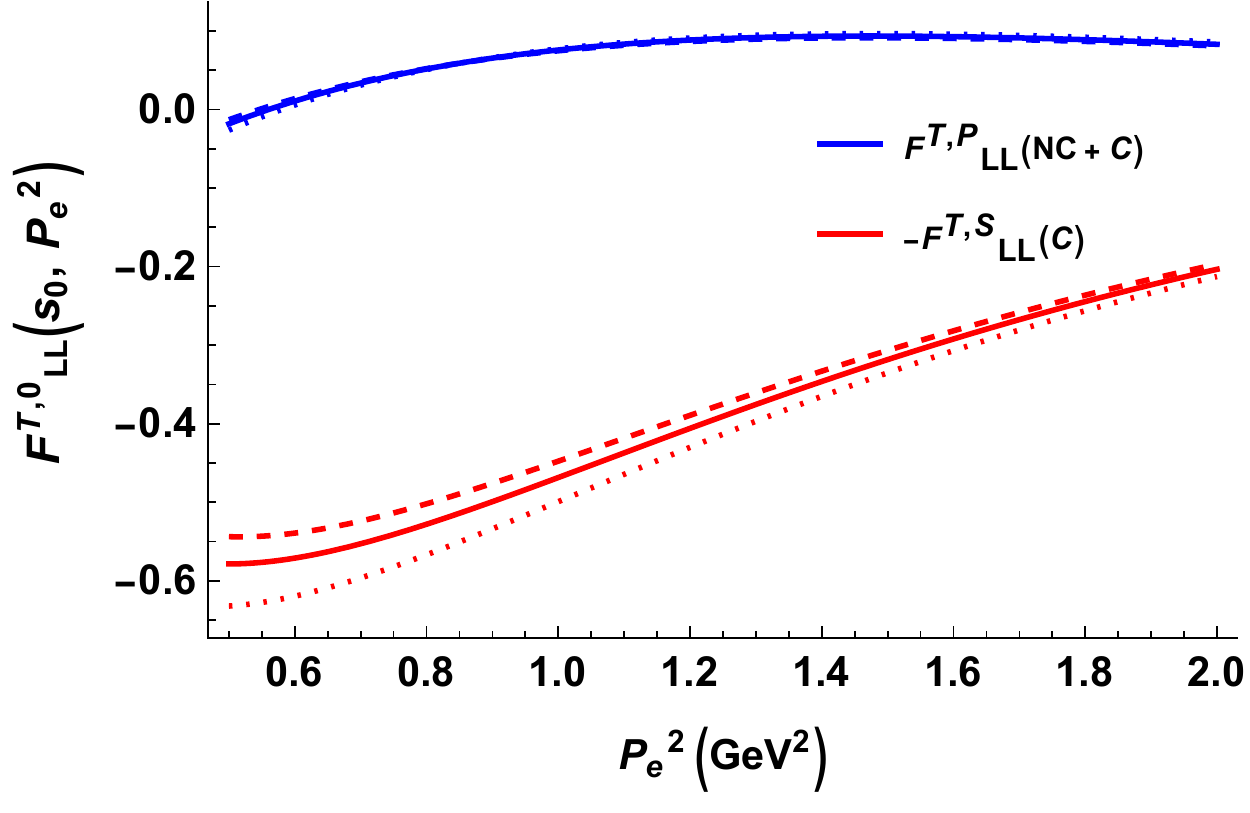}
   \end{subfigure}
   \begin{subfigure}{0.45\textwidth}
    \centering
    \includegraphics[width=0.8\linewidth]{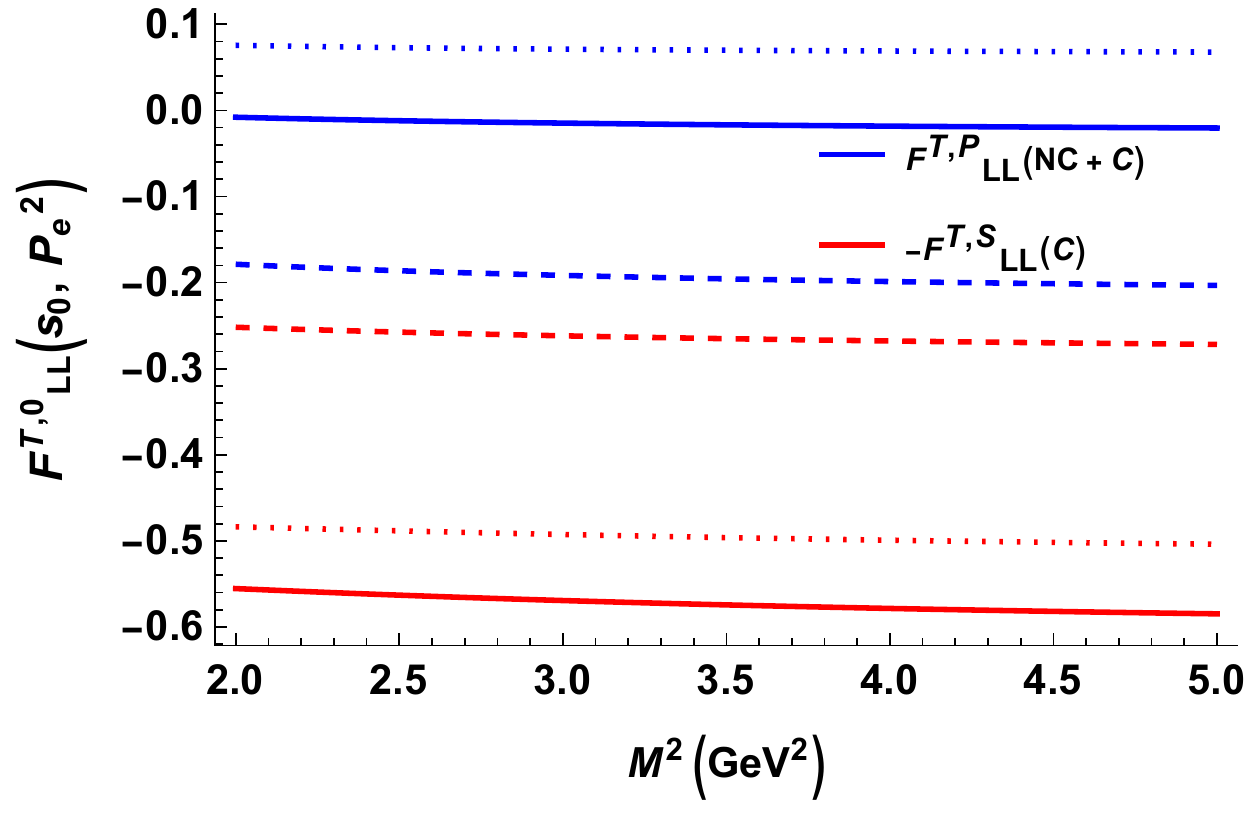}
   \end{subfigure}
   \begin{subfigure}{0.45\textwidth}
    \centering
    \includegraphics[width=0.8\linewidth]{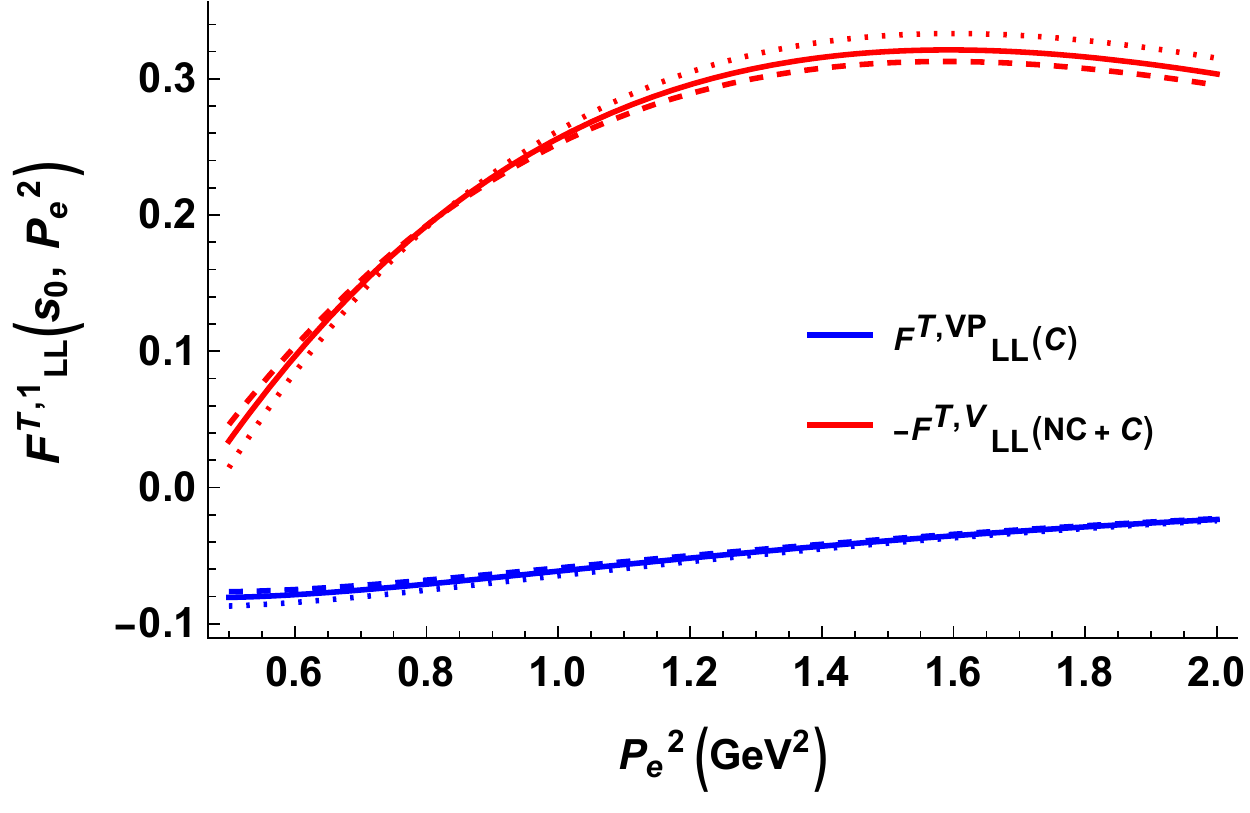}
   \end{subfigure}
   \begin{subfigure}{0.45\textwidth}
    \centering
    \includegraphics[width=0.8\linewidth]{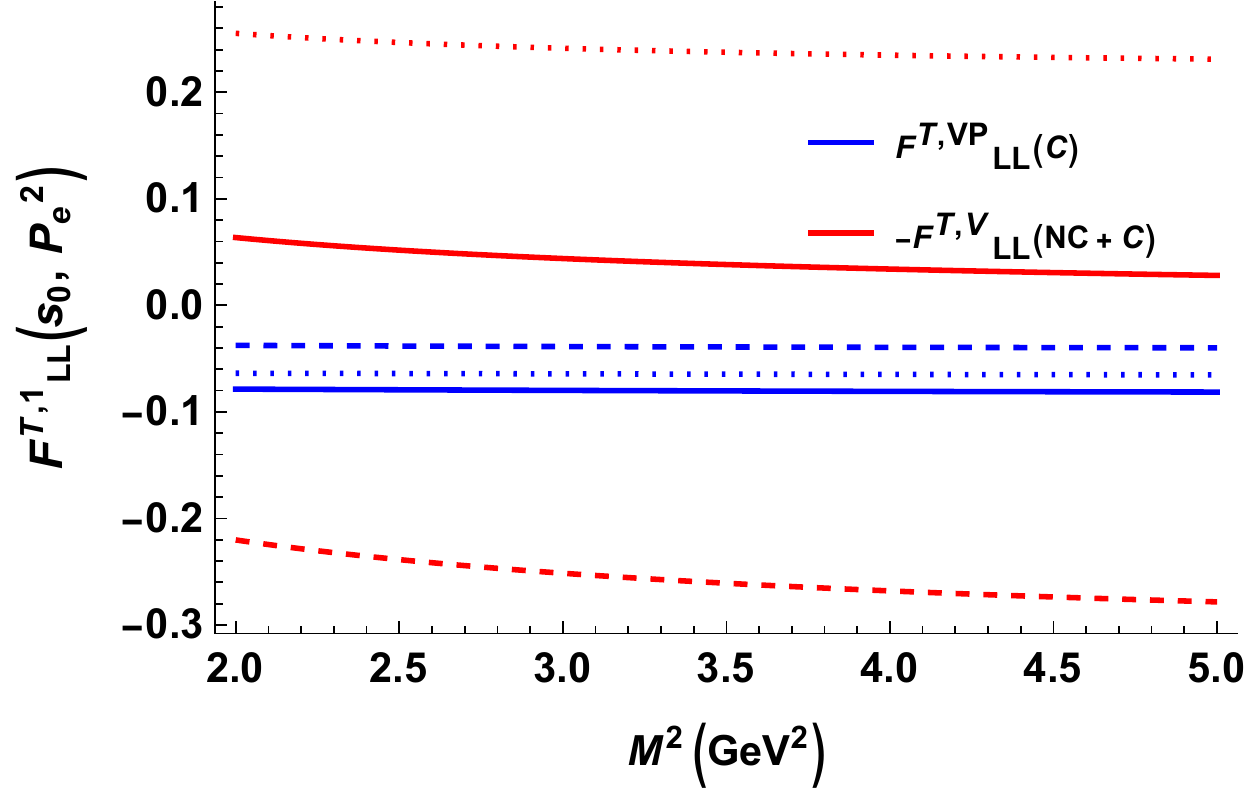}
   \end{subfigure}
    \caption{Same as Fig.(1) but for interpolation current, $\chi_{LA}$.}
    \label{LLt0}
\end{figure}

\begin{figure}[h]
\centering
   \begin{subfigure}{0.45\textwidth}
    \centering
    \includegraphics[width=0.8\linewidth]{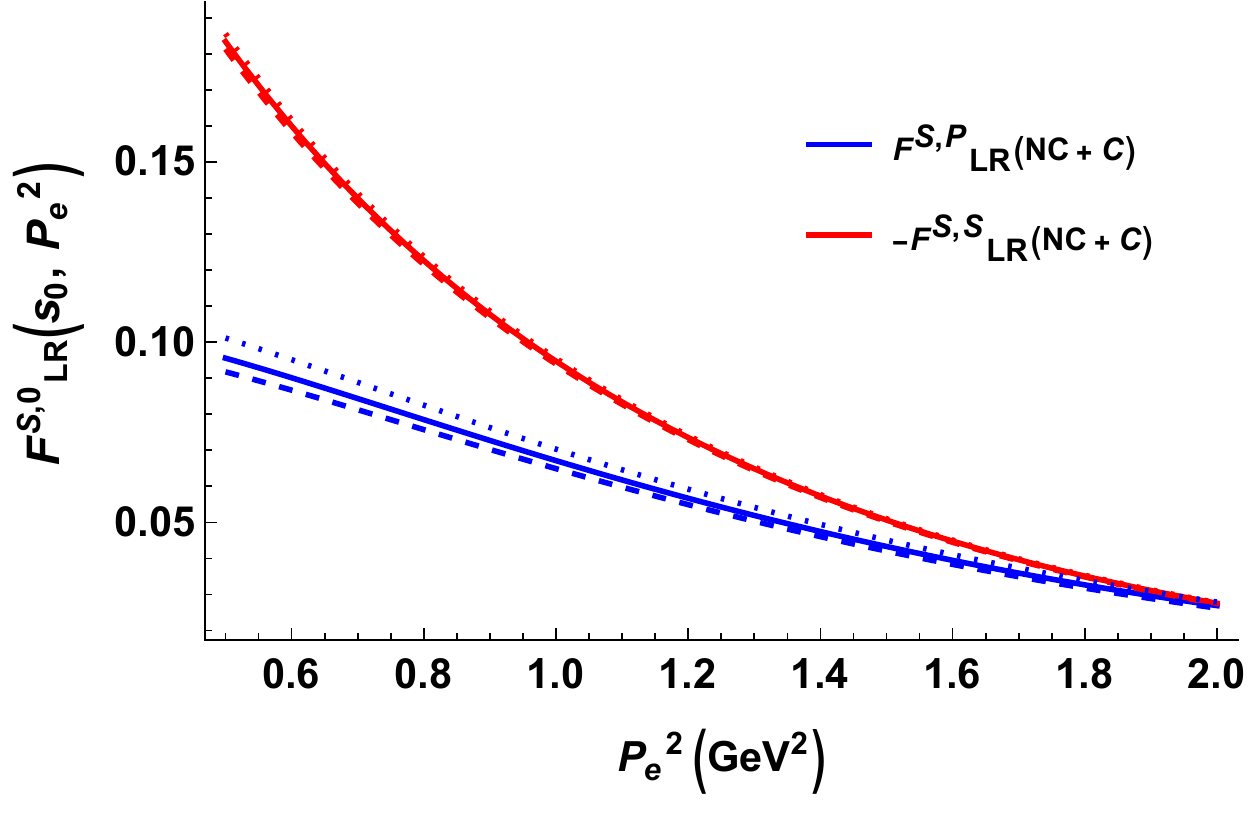}
   \end{subfigure}
   \begin{subfigure}{0.45\textwidth}
    \centering
    \includegraphics[width=0.8\linewidth]{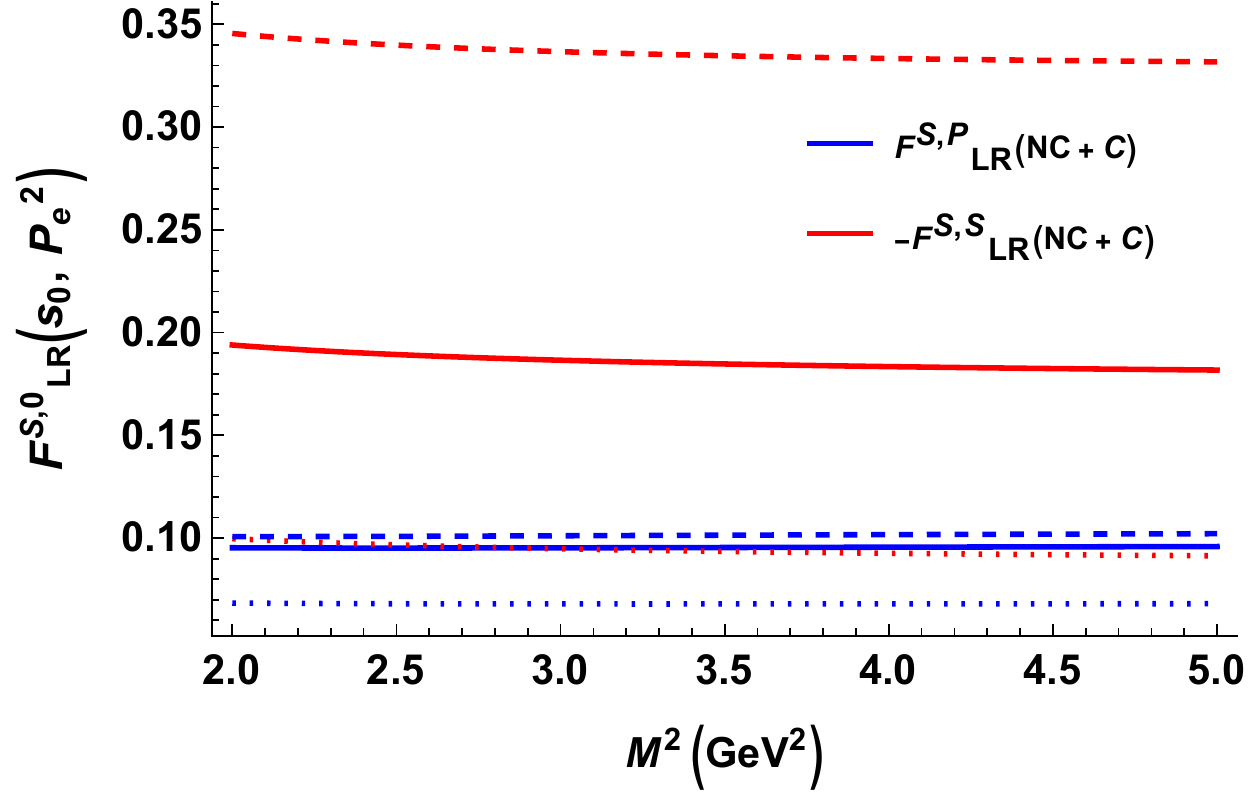}
   \end{subfigure}
   \begin{subfigure}{0.45\textwidth}
    \centering
    \includegraphics[width=0.8\linewidth]{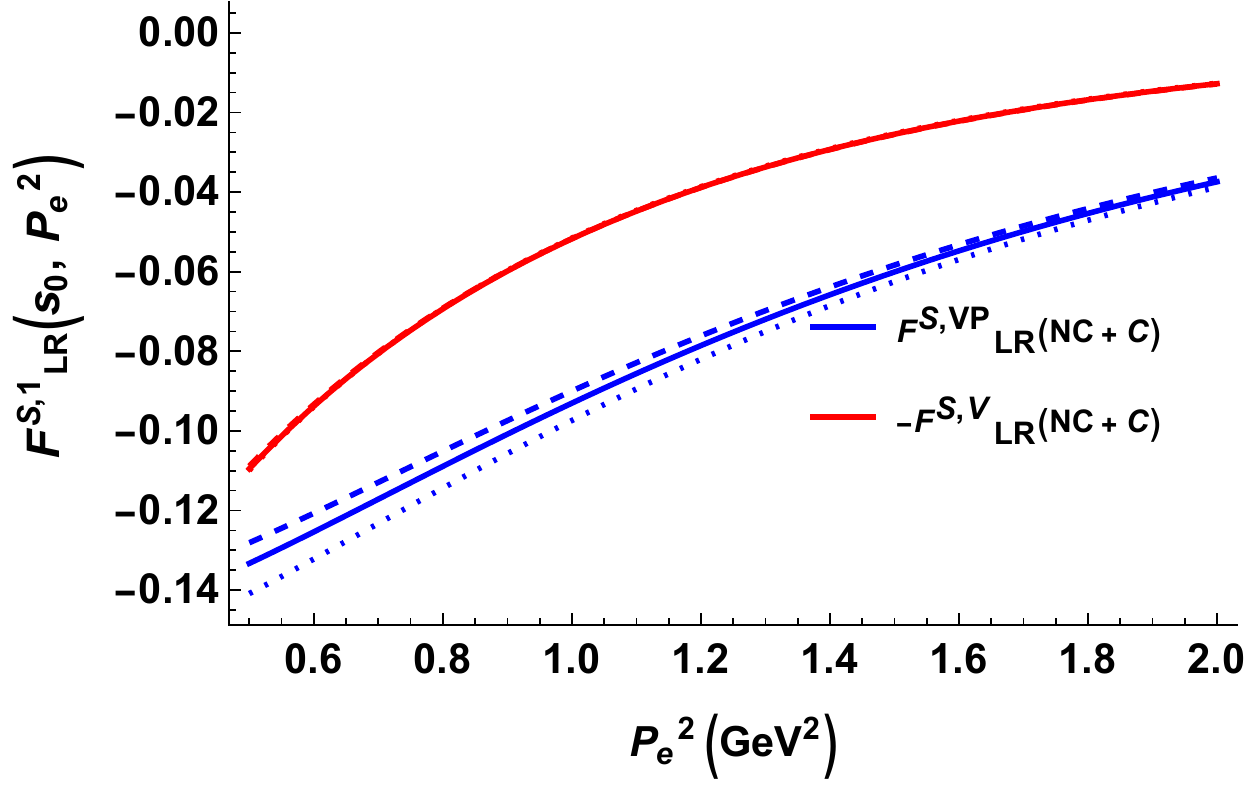}
   \end{subfigure}
   \begin{subfigure}{0.45\textwidth}
    \centering
    \includegraphics[width=0.8\linewidth]{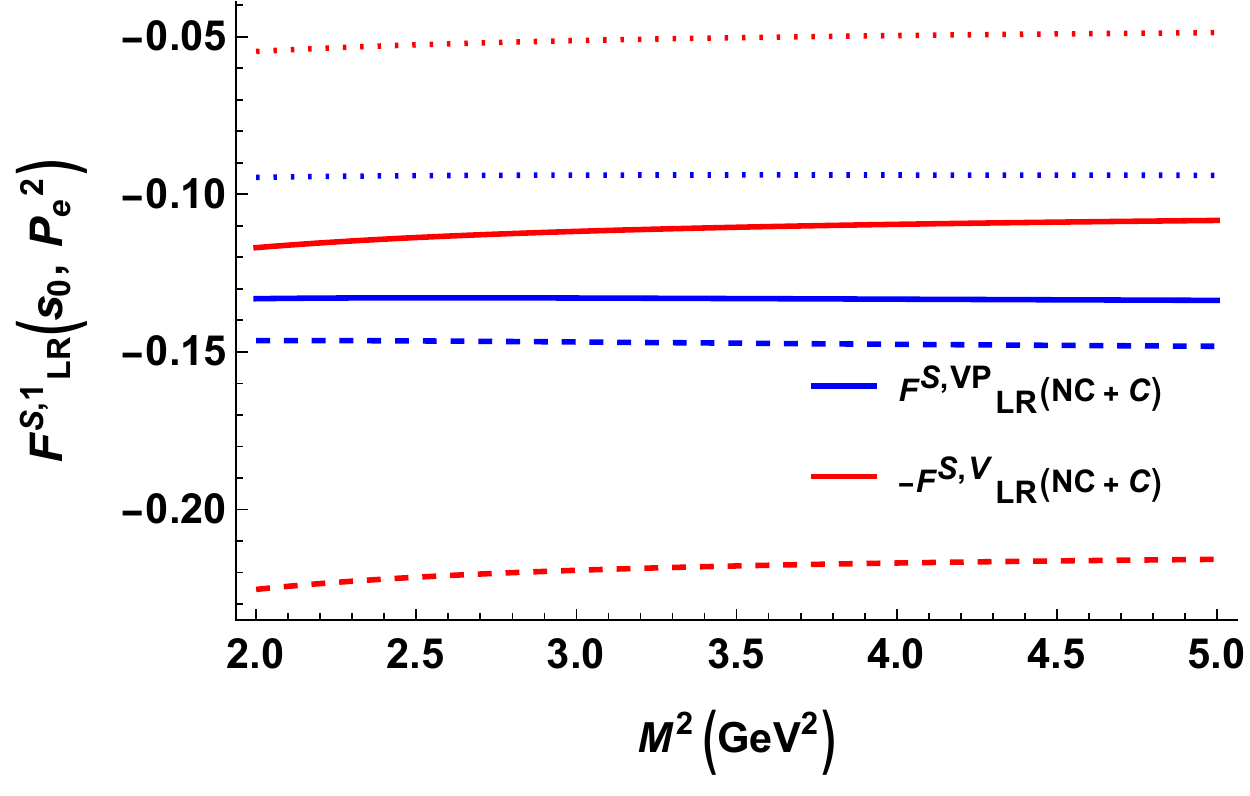}
   \end{subfigure}
   \begin{subfigure}{0.45\textwidth}
    \centering
    \includegraphics[width=0.8\linewidth]{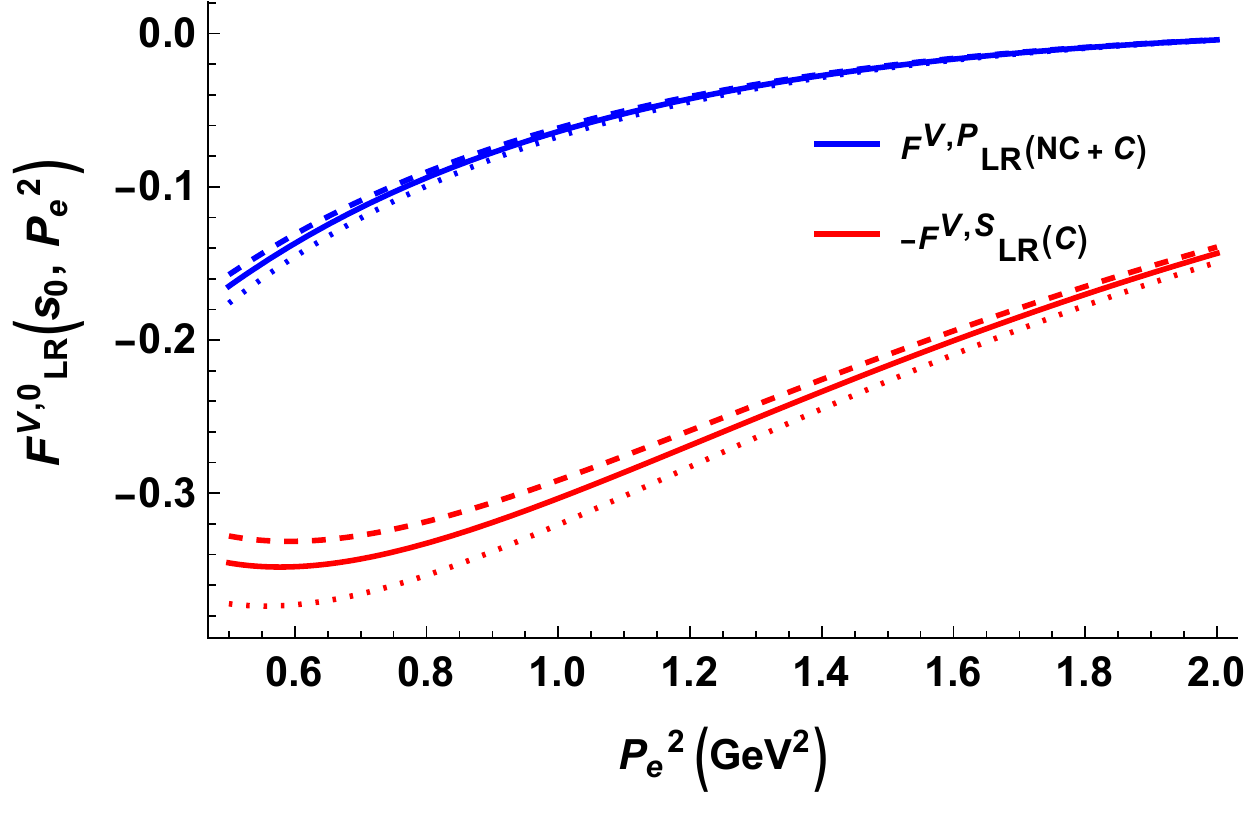}
   \end{subfigure}
   \begin{subfigure}{0.45\textwidth}
    \centering
    \includegraphics[width=0.8\linewidth]{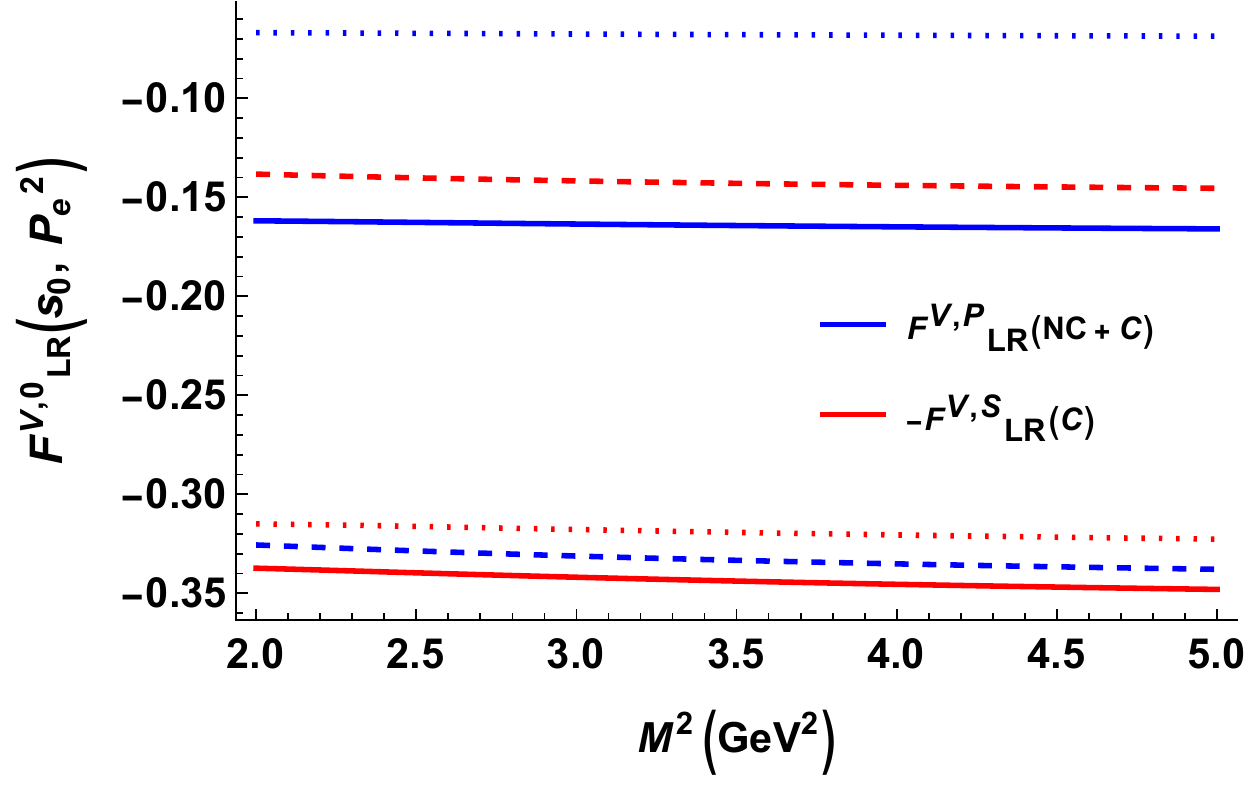}
   \end{subfigure}
   \begin{subfigure}{0.45\textwidth}
    \centering
    \includegraphics[width=0.8\linewidth]{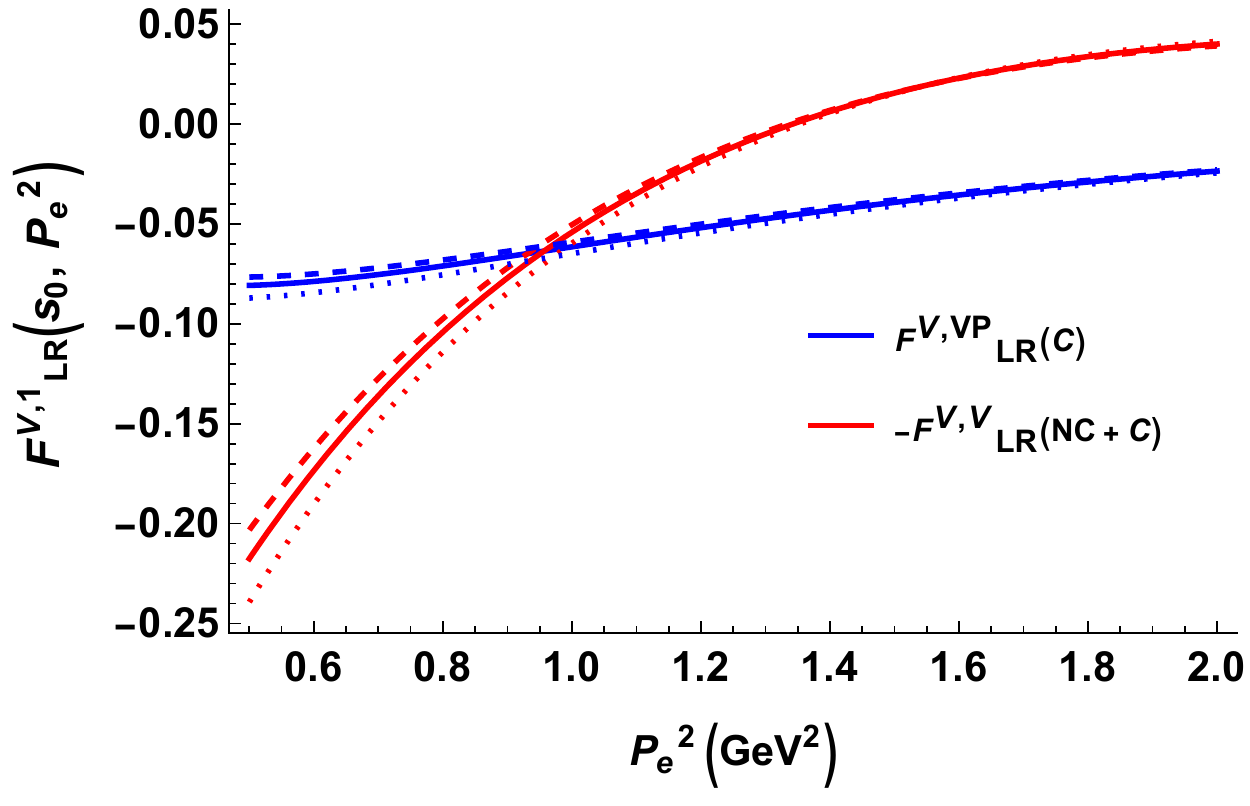}
   \end{subfigure}
   \begin{subfigure}{0.45\textwidth}
    \centering
    \includegraphics[width=0.8\linewidth]{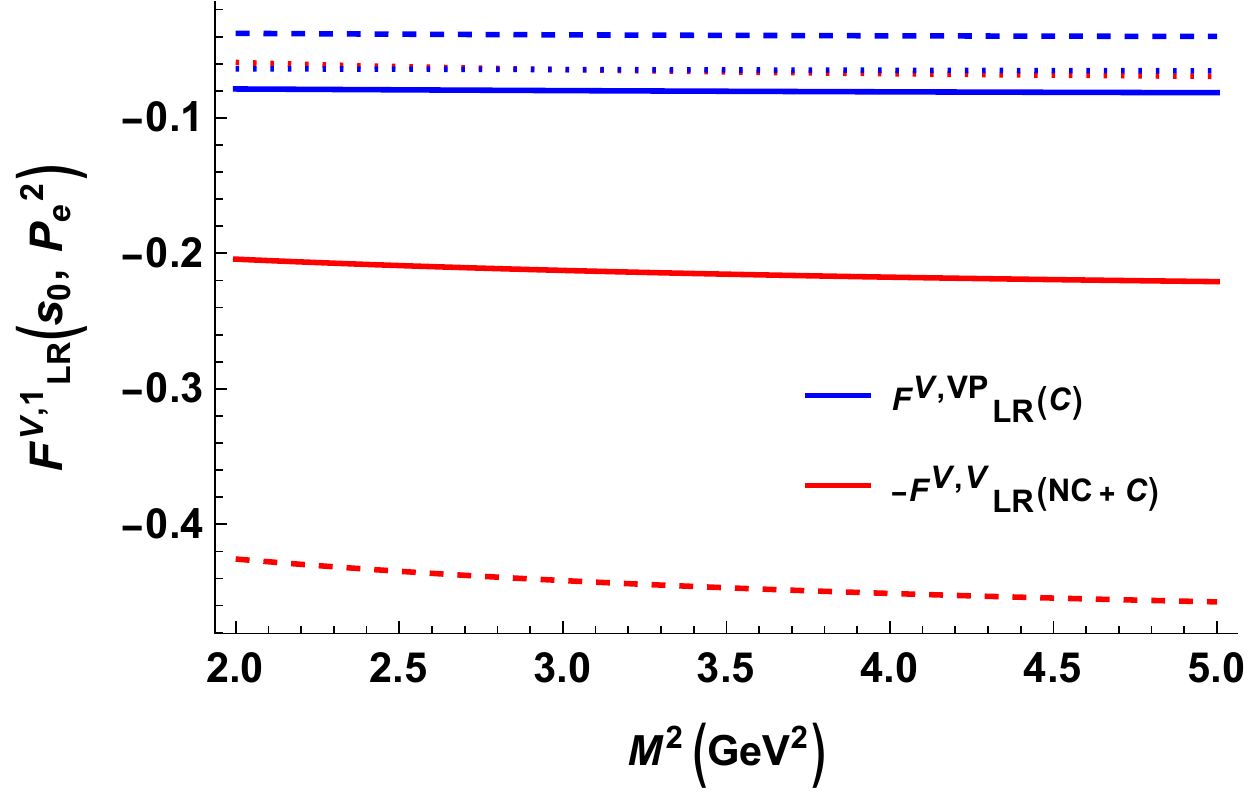}
   \end{subfigure}
    \caption{Same as Fig.(3) for $F_{LR}^{A,n}$ extracted from $F_{LR}^{A,r}$.}
    \label{LRt0}
\end{figure}

\section{Conclusions and Discussion}
\label{CD}
The form factors involved in the BNV process $D^0\rightarrow \bar p e^+$ are calculated in the framework of light cone sum rules using the distribution amplitudes of D-meson. We have found that this process involves 12 independent FFs, each of which can be extracted from two $F_{\Gamma\Gamma'}^{A,r}$. The extraction from the two does not match completely. There might be two reasons for that. Firstly, some of $F_{\Gamma\Gamma'}^{A,r}$ consists of only condensate contribution while others have both condensate and non-condensate contributions. The case where both the combinations have condensate as well as non-condensate contributions, like for $F_{LL}^{S,0}$ and $F_{LL}^{S,1}$ for $\chi_{LA}$ case (see Fig.(\ref{LLt0})), the extractions from the two combinations are close to each other. Secondly, as the LCSR predictions are more trustworthy at large $P_e^2$, the extractions from different combinations might be different at low $P_e^2$. At large $P_e^2$, they seem to be approaching each other. The similar analysis using different methods like lattice QCD will shed some light on the validity of different extractions. The higher order effects are also required to be included to get better clarity in future works. Also, the errors are found to be very large which are mainly dominated by the uncertainty in the model input parameter, $w_0$ in the LCDAs of D-meson. Better understanding of these LCDAs are required to required to get better understanding of these FFs. In this work we have taken a first step in computing FFs for baryon number violating decays of the heavy charm meson: $D^0\rightarrow \bar p e^+$. Other modes like $D^0\rightarrow\bar \Lambda e^+$ can be studied straightforwardly using the same method. As experimental searches improve, it is required to have first estimates of these non-perturbative inputs. Even though FFs extracted from two combinations do not match in some cases, they are numerically within the error bars of each other and thus provide a reasonable estimate. These can be used in a specific model framework where $c_{\Gamma\Gamma'}^A$'s are known functions of heavy particle masses and couplings to obtain the bounds on the parameters of the theory.
\bigskip
\acknowledgments
I thank Namit Mahajan for fruitful discussions. 

\appendix
\section{Useful definitions and integrals}
\label{A}
\subsection{Proton interpolation current}
As discussed in Section-3, the interpolation current, $\chi(x)$ for the proton state is not unique. The general form for $\chi(x)$ is adopted in Eq.(3.1) as
\begin{equation}
    \chi(x) = \chi_1(x)+t\chi_2(x)
\end{equation}
with $\chi_1(x)$ and $\chi_2(x)$ defined in Eq.(3.2) as
\begin{equation}
     \chi_1(x)= \epsilon^{lmn}\left(u^T_l(x) C \gamma_5 d_{m}(x)\right)u_n(x), \hspace {1cm}\chi_2(x)= \epsilon^{lmn}\left(u^T_l(x) C  d_{m}(x)\right)\gamma_5u_n(x)
\end{equation}
In general, the interpolation current used in LCSR calculation is given as
\begin{align}
    \chi_{IO}(x) &= 2\left(\chi_2(x)-\chi_1(x)\right) \nonumber \\
    & = \epsilon^{lmn} \left(u^{Tl}(x)C\gamma_\mu u^m(x)\right)\gamma_5 \gamma^\mu d^n(x)
\end{align}
such that, $\langle\bar p (p_p)|\chi_{IO}|0\rangle = m_p \lambda_{p1} v_p(p_p)$. This current is popularly known as the Ioffe current. This current can be obtained by taking $t=-1$ and multiplying $\chi(x)$ by -2 i.e. 
\begin{equation}
   \chi_{IO} (x)=-2 \chi(x)\hspace{1cm} \text{with  } t=-1.
   \label{chiio}
\end{equation}
However, the usual form of interpolation current used for lattice QCD computations, $\chi_{LA}(x)$ is given by $\chi(x)$ with $t=0$ i.e. 
\begin{equation}
    \chi_{LA}(x) = \chi(x) \hspace{1cm} \text{with  } t=0
    \label{chila}
\end{equation}
such that $\langle\bar p (p_p)|\chi_{LA}|0\rangle = m_p \lambda_{p2} v_p(p_p)$.
\subsection{Useful Integrals}
In this section we collect all the useful integrals required for the derivation of the sum rule for the FFs involved or simply the calculation of the correlation function in QCD. The general formula for the integrals in D-dimension which usually appear in the sum rule calculations is \cite{Khodjamirian:2020btr}
\begin{equation}
    \int d^D x e^{i p.x}\frac{1}{\left(x^2\right)^n}=\left(-i\right)\left(-1\right)^n 2^{(D-2n)}\pi^{D/2}\left(-p^2\right)^{n-D/2}\frac{\Gamma\left(D/2-n\right)}{\Gamma(n)}
\end{equation}
for $n\geq1$ ,$p^2<0$. The desired integrals for the present case can be obtained by differentiating it with respect to the four momentum $p_\alpha$. 
\begin{align}
    & \int d^4x \hspace{0.1cm} e^{i p.x}\frac{x_\alpha}{x^6} =\frac{-\pi^2}{4} p_\alpha \text{ln}(-p^2), \hspace{2cm} \int d^4x\hspace{0.1cm}  e^{i p.x}\frac{1}{x^6} =\frac{-i\pi^2}{8} p^2 \text{ln}(-p^2) \nonumber \\ &  \int d^4x \hspace{0.1cm} e^{i p.x}\frac{x_\alpha}{x^4} = 2\pi^2 \frac{p_\alpha}{p^2}, \hspace{3.2cm} \int d^4x\hspace{0.1cm}  e^{i p.x} \frac{1}{x^2}= \frac{-4i\pi^2}{p^2} \nonumber \\
    & \int d^4x e^{ipx}\frac{x_\alpha x_\beta}{x^8} =\frac{-i\pi^2}{48} \left(p^2g_{\alpha\beta}+2p_\alpha p_\beta\right) \text{ln}(-p^2) 
\end{align}
The divergent terms, proportional to $p^2$ are omitted here as they vanishes upon Borel transformation.
\section{Correlation Functions}
\label{C}
In this section we collect all the correlation functions calculated in QCD for different combinations of $\Gamma$, $\Gamma'$ and $A$ with 
\begin{align}
    P^2 &= (p_e+wv)^2 = ((w+m_D)v-p_p)^2\nonumber \\
    &= w(w+m_D)-\frac{ws}{m_D}  - \left(\frac{w+m_D}{m_D} \right)P_e^2
\end{align}
where, $s=p_p^2$ and $P_e^2=-p_e^2$. Also, as $v^2=1$ 
\begin{align}
    (v.P)&= -(v.p_p-(w+m_D))\nonumber \\
    &=  \frac{2w+m_D}{2} - \frac{s+P_e^2}{2m_D}
\end{align}
\subsection{Case-1: $P_\Gamma = P_\Gamma'= P_L$}
\begin{itemize}
    \item For $\Gamma^A=1$
    \begin{align}
        \Pi_{LL}^{S,S}&=\frac{f_Dm_D}{8}\int_0^\infty dw \left[\frac{(t-1)}{4\pi^2} \left\{(w+m_D)\Phi_{\pm}^D(w)+P^2\phi_+^D(w)\right\}\text{ln}(-P^2)\nonumber \right.\\ &+ \left. \frac{\langle\bar q q\rangle (t-1)}{3}\left\{\frac{(w+m_D)\phi_+^D(w)}{P^2}+\frac{\Phi_{\pm}^D(w)}{P^2}\right\}\right]
    \end{align}
     \begin{align}
        \Pi_{LL}^{S,V}&=\frac{f_Dm_D}{8}\int_0^\infty dw \left[\frac{3(t+1)}{8\pi^2} \left\{(w+m_D)\Phi_{\pm}^D(w)+P^2\phi_+^D(w)\right\}\text{ln}(-P^2)\nonumber \right.\\ &- \left. \frac{\langle\bar q q\rangle (t-1)}{3}\left\{\frac{(w+m_D)\phi_+^D(w)}{P^2}+\frac{\Phi_{\pm}^D(w)}{P^2}\right\}\right]
    \end{align}
     \begin{align}
        \Pi_{LL}^{S,P}&=-m_p\frac{f_Dm_D}{8}\int_0^\infty dw \left[\frac{3(t+1)}{8\pi^2} \Phi_{\pm}^D(w)\text{ln}(-P^2)- \frac{\langle\bar q q\rangle (t-1)}{3}\frac{\phi_+^D(w)}{P^2}\right]
    \end{align}
    \begin{align}
        \Pi_{LL}^{S,VP}&=-m_p\frac{f_Dm_D}{8}\int_0^\infty dw \left[\frac{(t-1)}{4\pi^2} \Phi_{\pm}^D(w)\text{ln}(-P^2)+ \frac{\langle\bar q q\rangle (t-1)}{3}\frac{\phi_+^D(w)}{P^2}\right]
    \end{align}
    \item For $\Gamma^A=\gamma_\mu$
    \begin{align}
        \Pi_{LL}^{V,S}&=\frac{f_Dm_D}{4}\int_0^\infty dw \left[\frac{(t-1)}{8\pi^2} \left\{(w+m_D)\Phi_{\pm}^D(w)+P^2\phi_+^D(w)\right\}\text{ln}(-P^2)\nonumber \right.\\ &+ \left. \frac{\langle\bar q q\rangle}{3}\left\{\frac{(t-1)\Phi_\pm^D(w)}{P^2}+\frac{(3+t)(w+m_D)\phi_{+}^D(w)}{P^2}-\frac{4 (v.P) \phi_+^D(w)}{P^2}\right\}\right]
    \end{align}
     \begin{align}
        \Pi_{LL}^{V,V}&=-\frac{f_Dm_D \langle\bar q q\rangle}{12}\int_0^\infty dw \left[\frac{(3+t)\Phi_{\pm}^D(w)}{P^2}-\frac{(w+m_D)}{P^2}\left\{2\phi_-^D(w)-(t+1)\phi_+^D(w)\right\}\right]
    \end{align}
     \begin{align}
        \Pi_{LL}^{V,P}&=-m_p\frac{f_Dm_D \langle\bar q q\rangle}{12}\int_0^\infty dw \left[\frac{1}{P^2}\left\{2\phi_-^D(w)-(t+1)\phi_+^D(w)\right\}\right]
    \end{align} 
    \begin{align}
        \Pi_{LL}^{V,VP}&=-m_p\frac{f_Dm_D}{4}\int_0^\infty dw \left[\frac{(t-1)}{8\pi^2} \Phi_{\pm}^D(w)\text{ln}(-P^2)+ \frac{\langle\bar q q\rangle (t+3) }{3}\frac{\phi_+^D(w)}{P^2}\right]
    \end{align}
    \item For $\Gamma^A=\sigma_{\mu\nu}$
    \begin{align}
        \Pi_{LL}^{T,S}&=\frac{f_Dm_D \langle\bar q q\rangle (t-1)}{6}\int_0^\infty dw \left[\frac{1}{P^2}\left\{4\phi_+^D(w)(v.P-(w+m_D))+3\Phi_\pm^D(w)\right\}\right]
    \end{align}
     \begin{align}
        \Pi_{LL}^{T,V}&=\frac{f_Dm_D}{2}\int_0^\infty dw \left[\frac{(t+1)}{4\pi^2} \left\{\frac{\phi_{+}^D(w)}{6}\left(4(w+m_D)(v.P)-P^2\right)+\frac{3}{2}\Phi_\pm^D(w)(w+m_D)\right\}\text{ln}(-P^2)\nonumber \right.\\ &- \left. \frac{\langle\bar q q\rangle (t-1)}{3}\left\{\frac{(w+m_D)(\phi_+^D(w)+2\phi_-^D(w))}{P^2}-\frac{\Phi_{\pm}^D(w)}{P^2}\right\}\right]
    \end{align}
     \begin{align}
        \Pi_{LL}^{T,P}&=-m_p\frac{f_Dm_D}{2}\int_0^\infty dw \left[\frac{(t+1)}{4\pi^2} \left\{\frac{3}{2}\Phi_{\pm}^D(w)+\frac{2 (v.P)}{3}\phi_+^D(w)\right\}\text{ln}(-P^2)\nonumber \right.\\ &- \left. \frac{\langle\bar q q\rangle (t-1)}{3}\left\{\frac{(\phi_+^D(w)+2\phi_-^D(w)}{P^2}\right\}\right]
    \end{align} \begin{align}
        \Pi_{LL}^{T,VP}&=m_p\frac{f_Dm_D \langle\bar q q\rangle (t-1)}{6}\int_0^\infty dw \left[\frac{\phi_+^D(w)}{P^2}\right]
    \end{align}
    
\end{itemize}
\subsection{Case-2: $P_\Gamma = P_L$ and $ P_\Gamma'= P_R$}
\begin{itemize}
    \item For $\Gamma^A=1$
    \begin{align}
        \Pi_{LR}^{S,S}&=\frac{f_Dm_D}{8}\int_0^\infty dw \left[\frac{3(t+1)}{8\pi^2} \left\{(w+m_D)\Phi_{\pm}^D(w)+P^2\phi_+^D(w)\right\}\text{ln}(-P^2)\nonumber \right.\\ &- \left. \frac{\langle\bar q q\rangle (t-1)}{3}\left\{\frac{(w+m_D)\phi_+^D(w)}{P^2}+\frac{\Phi_{\pm}^D(w)}{P^2}\right\}\right]
    \end{align}
    \begin{align}
        \Pi_{LR}^{S,V}&=\frac{f_Dm_D}{8}\int_0^\infty dw \left[\frac{(t-1)}{4\pi^2} \left\{(w+m_D)\Phi_{\pm}^D(w)+P^2\phi_+^D(w)\right\}\text{ln}(-P^2)\nonumber \right.\\ &+ \left. \frac{\langle\bar q q\rangle (t-1)}{3}\left\{\frac{(w+m_D)\phi_+^D(w)}{P^2}+\frac{\Phi_{\pm}^D(w)}{P^2}\right\}\right]
    \end{align}
    \begin{align}
        \Pi_{LR}^{S,P}&=-m_p\frac{f_Dm_D}{8}\int_0^\infty dw \left[\frac{(t-1)}{4\pi^2} \Phi_{\pm}^D(w)\text{ln}(-P^2)+ \frac{\langle\bar q q\rangle (t-1)}{3}\frac{\phi_+^D(w)}{P^2}\right]
    \end{align}
    \begin{align}
        \Pi_{LR}^{S,VP}&=-m_p\frac{f_Dm_D}{8}\int_0^\infty dw \left[\frac{3(t+1)}{8\pi^2} \Phi_{\pm}^D(w)\text{ln}(-P^2)- \frac{\langle\bar q q\rangle (t-1)}{3}\frac{\phi_+^D(w)}{P^2}\right]
    \end{align}
     \item For $\Gamma^A=\gamma_\mu$
    \begin{align}
        \Pi_{LR}^{V,S}&= \frac{f_Dm_D \langle\bar q q\rangle}{12}\int_0^\infty dw \left[\frac{\phi_+^D(w)}{P^2}\left\{(w+m_D)-(t+3)(v.P)\right\}-\frac{(t+2)\Phi_{\pm}^D(w)}{P^2}\right]
    \end{align}
    \begin{align}
        \Pi_{LR}^{V,V}&=-\frac{f_Dm_D}{4}\int_0^\infty dw \left[\frac{(t-1)}{4\pi^2} \left\{(w+m_D)\Phi_{\pm}^D(w)+\frac{\phi_+^D(w)}{6}\left(P^2+2(v.P)(w+m_D)\right)\right\}\text{ln}(-P^2)\nonumber \right.\\ &- \left. \frac{\langle\bar q q\rangle}{3}\left\{\frac{(w+m_D)}{P^2}\left(\phi_-^D(w)(t+3)+\phi_+^D(w)(t+1)\right)-\frac{\Phi_{\pm}^D(w)}{P^2}\right\}\right]
    \end{align}
    \begin{align}
        \Pi_{LR}^{V,P}&=m_p\frac{f_Dm_D}{4}\int_0^\infty dw \left[\frac{(t-1)}{4\pi^2} \left\{\frac{\phi_{+}^D(w)(v.P)}{3}+\Phi_\pm^D(w)\right\}\text{ln}(-P^2)\nonumber \right.\\ &- \left. \frac{\langle\bar q q\rangle}{3 P^2}\left\{\phi_-^D(w)(t+3)+\phi_+^D(w)(t+1)\right\}\right]
    \end{align}
    \begin{align}
        \Pi_{LR}^{V,VP}&=-m_p \frac{f_Dm_D\langle\bar q q\rangle}{6}\int_0^\infty dw \frac{\phi_+^D(w)}{P^2}
    \end{align}
     \item For $\Gamma^A=\sigma_{\mu\nu}$:
    All the correlation functions are zero.
\end{itemize}
Extracting the imaginary of part of these correlation functions is rather easy as,
\begin{align}
    & \text{Im}\left(\frac{1}{P^2-m^2}\right) = -\pi \delta(P^2-m^2)\nonumber \\
    & \text{ln}(-x) =\text{ ln}|x|-i\pi \theta(x)
\end{align}
where, $m $ is some mass scale and is zero our case. Hence, to get the imaginary part, just replace $\text{ln}(-P^2)$ by $-\pi \theta(P^2)$ and $\frac{1}{P^2}$ by $-\pi\delta(P^2)$ in the correlation functions.
\section{Numerical Values of parameters used}
\label{D}
In this appendix, we collect all the numerical values of the parameters used during numerical analysis\footnote{The decay constant for $D^0$ meson is not known. We have used the decay constant for $D^+$ meson here.} .
\begin{center}
    \begin{tabular}{|c||c|c|c|}
    \hline
 \textbf{S.No.}&  \textbf{ Parameter} & \textbf{Value Used} & \textbf{Reference}\\
 \hline\hline
  \textbf{1.}& Proton mass ($m_p$)& 0.938 GeV &  \cite{ParticleDataGroup:2020ssz}\\
    \hline
     \textbf{2.} &Quark condensate ($\left<\bar q q\right>$) & $-((256\pm2)\text{MeV})^3$ & \cite{Haisch:2021hvj}\\
    \hline
     \textbf{3.} & D-meson decay constant, $f_D$& $(0.212\pm 0.001) \text{ GeV}^2$  & \cite{ParticleDataGroup:2020ssz} \\
     \hline
       \textbf{4.}& D-meson mass, $m_D$ & $1.864 \text{ GeV}$  & \cite{ParticleDataGroup:2020ssz}\\
       \hline
    \textbf{5.} & $\lambda_{p1}$ & $(-0.027\pm0.009) \text{ GeV}^3$ & \cite{Braun:2006hz} \\
    \hline
    \textbf{6.} & $\lambda_{p2}$ & $(-0.013\pm0.004)\text{ GeV}^3$ & \cite{Leinweber:1994nm}\\
    \hline
    \textbf{7.} & $w_0$ & $(0.45\pm0.3) \text{ GeV}$ & \cite{Feldmann:2017izn}\\
    \hline
\end{tabular}
\end{center}

\bibliography{Dtope}{}
\bibliographystyle{unsrt}
\end{document}